\documentclass{aa}
\usepackage{graphicx}
\usepackage{natbib}
\begin{document}

   \title{Stellar Populations in HII Galaxies}

   \author{P. Westera
          \inst{1}
          \and
          F. Cuisinier\inst{1}
          \and
          E. Telles\inst{2}
          \and
          C. Kehrig\inst{2}
          }

   \offprints{F. Cuisinier}

   \institute{GEMAC, Observat\'{o}rio do Valongo/UFRJ,
             Ladeira do Pedro Ant\^{o}nio, 43, 20.080-090 Rio de Janeiro - RJ, Brazil\\
             \email{westera@ov.ufrj.br, francois@ov.ufrj.br}
         \and
             Observat\'{o}rio Nacional,
             Rua Jos\'{e} Cristino, 77, 20.921-400, Rio de Janeiro - RJ, Brazil\\
             \email{etelles@on.br, kehrig@on.br}
              }

   \date{Received <date>; accepted <date>}

   \abstract{We analyse the stellar content of a large number of HII
   galaxies from the continua and absorption features of their spectra
   using population synthesis methods, in order to gain information
   about the star formation histories of these objects.

   We find that all galaxies of our sample contain an old stellar
   population ($\geq 1$ Gyr) that dominates the stellar mass,
   and in a majority of these we also found evidence for an
   intermediate-age population $\geq 50$ Myr apart from the
   presently bursting, ionizing young generation $\leq 10^7$ yrs.

    }

   \maketitle

\section{Introduction}
\label{intro}

   HII galaxies are easily recognised from their prominent emission line
   spectra, very similar to those of HII regions (recombination lines of
   hydrogen and helium, as well as forbidden lines of elements like oxygen,
   neon, nitrogen, sulphur, and others, mainly in their first and second
   ionization stages).
   The continuum part of their spectra in the visible is much fainter,
   and mostly of stellar origin. 
   HII galaxies show very intense star formation, which is responsible
   for the ionization of the gas and the subsequent emission lines.
   In most cases,
   HII galaxies possess old populations as well,  that can generally
   be detected at a reasonable distance from their
   centers \citep{tellesterlevich}. 

   For dwarf galaxies, which constitute a significant fraction of HII
   galaxies, the gas content is however too low  for the star
   formation rate to have been sustained at the present level during
   their whole life. It is generally believed \citep[e.g.][]{searle} that
   they spend most of their lives in a quiescent phase, and that the
   intense star formation phases in which they are encountered now are very
   short.
   The old population that is generally detected would then be the
   result of the accumulation of former bursts. Some exceptions may exist,
   like IZw18, where some authors say that the young bursting population
   is the only one \citep[e.g.][]{papaderos},
   although others argue that a faint old underlying
   population does indeed exist, only detectable in the outskirts of the
   galaxy \citep[e.g.][]{ostlin,aloisi}, and may be the result of
   an early low level continuous phase of star formation, as proposed by
   \citet{legrand}. 

   The bursting mechanisms are still unknown, although various have
   been proposed, of internal - (e.g. interaction of hot and cold gas
   phases, like in \citet{tenoriotagle}, see also \citet{tellester})
   or external origin (e.g. tidal forces \citep{salzer}, ram pressure
   \citep{murakami}). The burst pattern is unclear as well: Does star
   formation occur simultaneously over the whole extent of the
   galaxy, or is the present star forming region formed by the
   accumulation of successive smaller star forming events,
   spread over the galaxy, over a short timescale?

   Although stellar population synthesis analyses cannot give any
   direct answer to the question of the origin of the bursts, they can
   be quite helpful in bringing some insight in the recent star
   formation history of these galaxies, and to a lesser extent in the
   older one, and thus contribute to answering our second question,
   e.g. what is the bursting pattern of HII galaxies? On this subject,
   we already have some evidence that the recent star
   formation is synchronised on scales of hundreds of parsecs rather
   than self-propagated. This is what Telles et al. found from
   a spatial-temporal analysis of optical and infrared surface
   photometry of a sample of HII galaxies \citep{telles,tellessampson}.
   It can also be seen from the H$_{\alpha}$ imaging atlas
   of blue compact dwarfs of \citet{gil}.
   Furthermore, stellar population synthesis allows us to put into
   evidence old populations (e.g. older than 1 Gyr), although their
   exact characteristics can be difficult to determine because of the
   mixture with much younger populations.

   In this paper, we use the continuum spectra of a large and
   homogeneous sample of HII galaxies to gain information
   about their (star formation) histories. Similar studies
   include the work done by \citet{raimann}, \citet{cidfernandes}, and
   \citet{kong}. \citet{raimann} use a star cluster spectral base to
   identify the stellar populations present in their sample. They
   find that HII galaxies are typically age-composite stellar systems,
   presenting important contributions from generations up to as old as
   500 Myr, and they detect a significant contribution of populations with
   ages older than 1 Gyr in two groups of HII galaxies. Cid Fernandes
   et al. use three absorption line strength indices and two continuum
   colours to place the spectra in an evolutionary diagram, whose axes
   carry the contribution of a young ionizing ($\leq 10$ Myr), an
   intermediate (100 Myr), and an old ($\geq 1$ Gyr) population to the
   total flux, as determined by comparing their indices with the same
   indices in synthetic spectra of these ages. They also find
   evidence for populations of all three age groups. With the same
   algorithm as Cid Fernandes et al., but using more absorption lines
   and the continuum fluxes at five wavelengths and a base of 35
   star-cluster spectra by \citet{bica}, \citet{kong} find that blue
   compact galaxies (many of which are HII galaxies) are typically
   age-composite stellar systems with stars from all three age groups
   contributing significantly to the 5870 \AA\ continuum emission of
   most galaxies in their sample.

   In this work we follow a similar approach: We define six spectral
   indices and determine for each spectrum the combination of the
   synthetic spectra of an old - and a young/intermediate stellar
   population, for which the indices of the empirical spectrum are best
   reproduced. The synthetic spectra are produced by two widely used
   evolutionary synthesis codes, GISSEL \citep{charlot_91, bruzual_93,
   bruzual_00} and STARBURST \citep{starburst}, one of which (GISSEL)
   implements the new BaSeL 3.1 stellar spectral library, (Basel
   standard Stellar Library 3.1) \citep{westera_02, westera_01}, which
   was calibrated using photometric data of globular cluster stars to
   improve the reproduction of the continua of the spectra of low
   metallicity stars. The goal of performing the analysis with two
   different libraries was on the one hand to check the reliability of
   the results, and on the other hand to find out through comparison of
   the results if the BaSeL 3.1 stellar library is able to reproduce the
   spectra of stars and populations beyond the parameter range of the
   objects used for its calibration, eg. if the library can also be used
   for young stars and populations. 

   The layout of this article is the following: In
   Sect.~\ref{spectra}, the catalog of spectra analysed in this work
   is presented. Sect.~\ref{method} gives a detailed description of
   the method we used to analyse the spectra. In Sect.~\ref{results},
   we discuss the accuracy of the method and its ability to give
   meaningful results. The main results are discussed in
   Sect.~\ref{distribution}.
   A summary and the main conclusions can be found in
   Sect.~\ref{conclusions}.

\section{The spectra}
\label{spectra}

   \begin{figure}
    \includegraphics[width=\columnwidth]{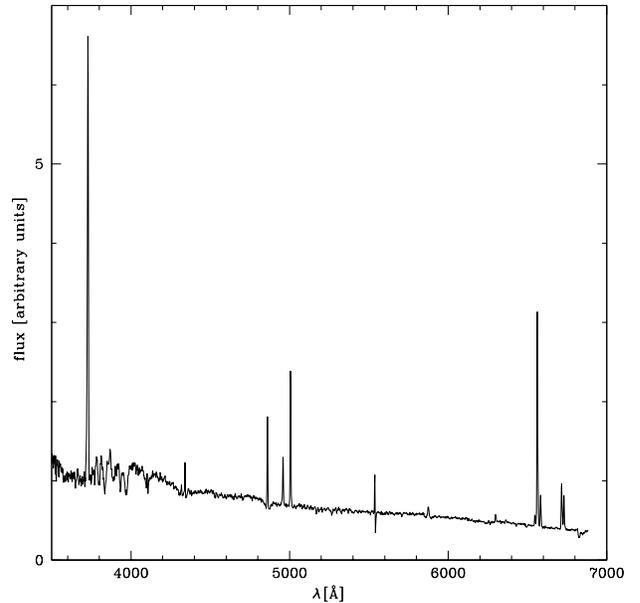}
    \caption{Example of an HII galaxy spectrum
        (UM 69, taken on August, the 18th, 1998).}
    \label{example}
   \end{figure}
   The data base of this work consists of a catalog of intermediate
   resolution ($R \sim 1000$) HII galaxy spectra in the range of 3700
   to 7500 \AA. It contains about 200 HII galaxy spectra observed with
   high signal-to-noise at the 1.52 m telescope at the European
   Southern Observatory (ESO) within the agreement between Brazil and
   ESO. All spectra were observed with the same instrumentation and
   reduced in a homogeneous way.

   The apertures were centered on star forming knots, and our
   work thus presents a bias towards young populations (as a matter of
   fact like all the HII galaxies spectroscopic surveys made up to
   now). The apertures we chose had the size of a few typical seeing
   widths, generally encompassing several stellar clusters. We
   therefore do not expect to have pure single stellar populations,
   but rather local mixes. Another important point is that a
   significant fraction of the galaxies were observed in several
   apertures, so we do not get only the
   most prominent knot, but also secondary ones (contrary to other
   spectroscopic studies), where the young population(s) is (are) less
   dominant in the integrated light. Table~\ref{indices} gives our
   galaxy sample and measured spectral indices as described below.
   Column 1 lists the names of the spectra, consisting of
   the names of the galaxies, and, in those cases where the apertures
   were centered on a secondary knot, an indication between brackets
   about the position of the aperture, where ``Cent'' stands for
   the centre of the
   galaxy, and the abbreviations ``N'', ``E'', ``S'', and ``W'' (and
   ``NE'' etc.) indicate that the aperture was positioned at knots at
   the North, East, South or West of the galaxy.
   Column 2 gives the type of the spectra (``in'': integrated spectra,
   ``re'': spectra of individual regions).
   For more details about the catalog, see \citet{kehrig}. 

   The spectra were corrected for redshift, and then for internal gas
   extinction.
   The latter was done by converting the absorption constants
   $C_{H_{\beta}}$ as determined by \citet{kehrig} into extinction
   constants $E(B-V)_{gas}$ using
   $E(B-V)_{gas}=0.665\times C_{H_{\beta}}$ \citep{leda}, then
   correcting for systematic differential extinction between the
   stellar populations and the gas employing
   $E(B-V)_{cont}=0.44\times E(B-V)_{gas}$ \citep{calzetti} (which
   leads to $E(B-V)_{cont}=0.2926\times C_{H_{\beta}}$), and finally
   dereddening the spectra using the extinction law of \citet{fluks}.

   Fig.~\ref{example} shows a typical corrected spectrum. Apart from
   the typical emission lines of an HII region, most spectra also show
   significant continuum contributions from a mixture of stellar
   populations. A closer examination of this continuum reveals that
   many of these spectra show signatures of young/intermediate
   populations (certain absorption lines, i. e. of hydrogen, and a
   blue continuum), as well as spectral properties of old populations
   (a 4000 \AA\ break, and absorption lines of heavy elements like
   Calcium, Iron, and Magnesium), which leads to the conclusion that
   many of the galaxies of the sample have undergone periods of star
   formation before the present ionizing massive stellar population.
   The goal of this work is to analyse the stellar content of these
   galaxies, and gain information about their star formation
   histories, using the continua and the most evident absorption
   features of their spectra.

\section{Method}
\label{method}
 
   \begin{figure*}
    \includegraphics[width=\textwidth]{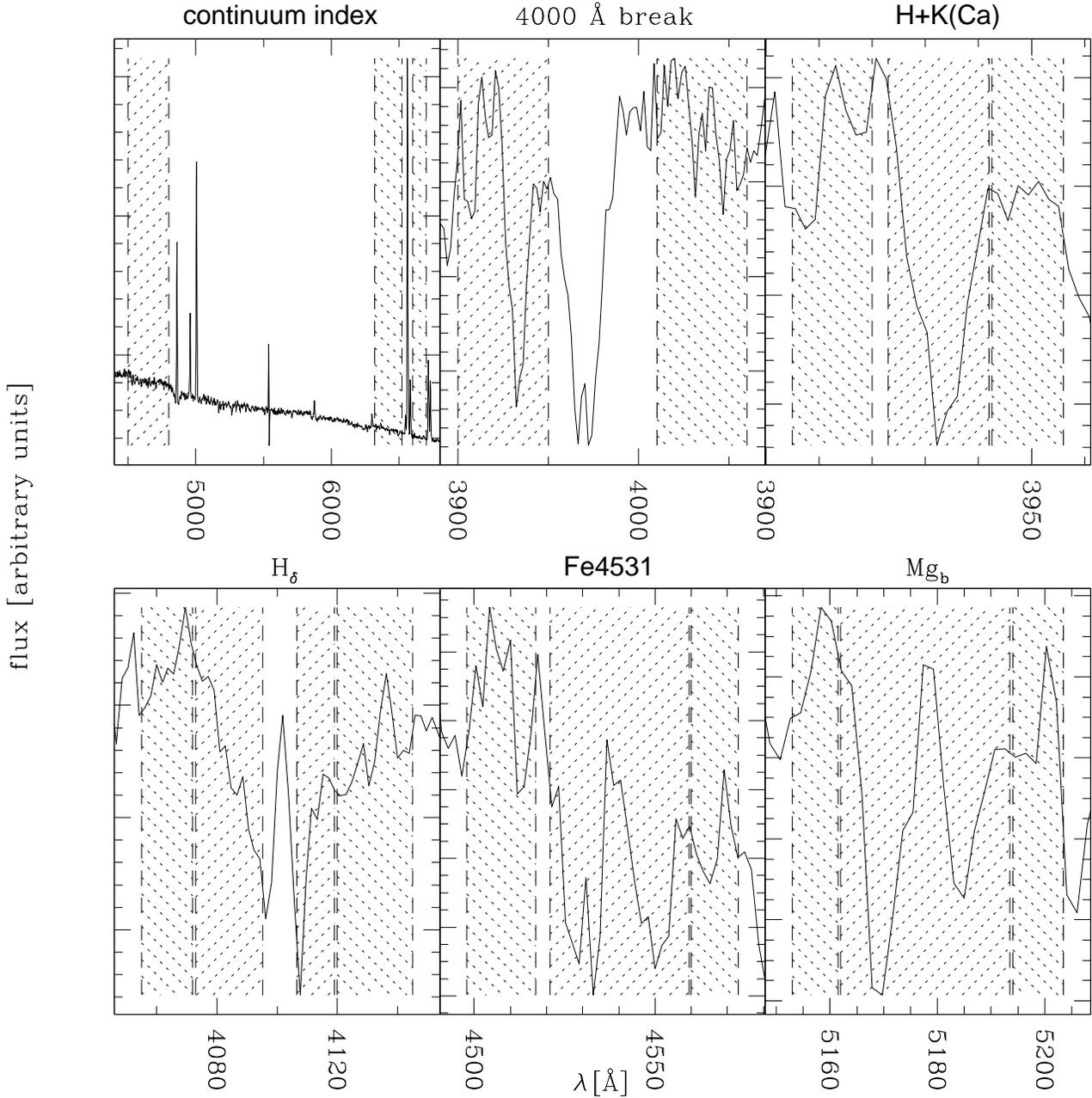}
    \caption{Illustration of the six indices using UM 69 (the spectrum
        shown in Fig.~\ref{example}) as an example. The dashed lines show
        the limits of the bands. The bands
        that enter with a positive sign into
        the calculation of the respective index are shaded with dotted lines
        running from the upper left to the lower right, and
        the bands that have a negative sign are shaded with dotted lines
        running from the upper right to the lower left. For a more detailed
        description of the indices, see the text.}
    \label{features}
   \end{figure*}
   We wish to characterise the population mixture in  our galaxies
   using the sensitivity of their spectra to the properties of these
   populations, e.g. abundances -- or metallicity -- and ages.
   Although we could directly fit modeled spectra to the observed ones,
   and evaluate the various population parameters (e.g. ages and
   metallicities) as the ones corresponding to the ``best-fitting'' model 
   spectrum, we prefer to proceed in a more conservative way, and use a set of
   pre-defined indices, chosen for their sensitivity to the parameters we
   want to derive. The fitting procedure is thus done to the indices, and
   not to the raw spectra. This presents the advantage of emphasising only
   useful features in the spectra,  giving a zero weight to parts that
   either are badly flux calibrated (e.g. wavelengths lower than
   ${\rm \lambda 3900 \AA}$), are contaminated by emission lines from
   the gas, or are very noisy, or simply provide redundant
   information. Because of the finite (and relatively low) number of
   indices, the maximum number of free parameters we adopt to
   characterise our galaxies is constrained in a straightforward manner.

   We  define six indices, relatively independent from each other, quantifying
   certain properties of the spectra, which are particularly suitable
   for identifying and characterising the stellar populations present in the galaxies
   (the shape of the continuum, the 4000 \AA\ break D(4000), and certain
   spectral (absorption) line strength indices).
   Most of the indices are inspired by the indices defined by Worthey et al.
   \citep{worthey,worthey_2} and carry the same names, and by analogy are
   defined in the following way:
   \begin{equation}
   {\rm Mag}=-2.5\log (F_{-}/F_{+})
   \end{equation}
   whereas:
   \begin{equation}
   F_{X}=\frac{1}{X}\int_{X} F_{\lambda} d\lambda
   \end{equation}
   Note that a wavelength range $X$ can consist of several disjoint
   wavelength ranges (i. e. the bands to the left and right of a spectral
   line). The wavelength ranges of the six indices are given in 
   Table~\ref{indexranges}, and illustrated in Fig.~\ref{features}.

   \begin{table}
   \begin{center}
      \caption{Wavelength ranges of the positive - and negative bands of the
      indices, with the associated errors.}
      \label{indexranges}
         \begin{tabular}{llll}
            \hline
 index &  $+$ range[\AA] &  $-$ range[\AA] & ${\rm \sigma_{index}}$ \\
            \hline
 continuum index & 6320-6520 & 4500-4800 &  0.23 \\
                 & 6600-6700 &  \\
 4000 \AA\ break & 4010-4060 & 3900-3950 &  0.03  \\
 H+K(Ca)         & 3905-3920 & 3923-3942 &  0.05  \\
                 & 3942.5-3956 &  \\
 H$_{\delta}$    & 4055-4072 & 4073-4095.25 &  0.04  \\
                 & 4120-4145 & 4106.5-4119  \\
 Fe4531          & 4498-4517 & 4521-4559.5  &  0.04 \\
                 & 4560-4573 &  \\
 Mg$_b$          & 5153-5161.5 & 5162-5193.5 & 0.05  \\
                 & 5194-5203.5 &  \\
            \hline
         \end{tabular}
   \end{center}
   \end{table}
   The continuum index is something like the $B-R$ colour with
   rectangular passbands. For some spectra, it was necessary to cut
   out a few tens of \AA\ due to contamination of the spectra by
   telluric skylines. For the other indices, the wavelength ranges
   may differ slightly from the
   ranges commonly used, since we optimised them for the spectra of
   our sample. The bands of the 4000 \AA\ break were reduced to a
   width of 50 \AA\ each because of the calibration uncertainties of
   the spectra downwards of $\lambda \sim 3900$ \AA\ affecting the $-$
   band and the presence of the strong H$_{\delta}$ emission line
   within the $+$ band of the traditional definition of this index. Of
   the H$_{\delta}$ line, only the wings are used, because the region
   around the central wavelength of this line is dominated by the
   emission line of the gas. The line strength indices are defined
   such that absorption goes to positive values and emission to
   negative ones. Since all our indices are absorption features, they
   are in general positive. However, as we do not make any correction
   for the inclination of the continuum, the wavelength localisation
   of the various bandpasses may sometimes induce negative values, as
   is for instance the case for H$_{\delta}$. We preferred not to make
   such corrections and to use indices instead of equivalent widths,
   because we prefer to use quantities that are more directly linked
   to the observed spectra, with a minimum number of in-between steps.
   The calculated indices of individual spectra are listed in
   Table~\ref{indices}.

\begin{figure*}
    \includegraphics[width=\textwidth]{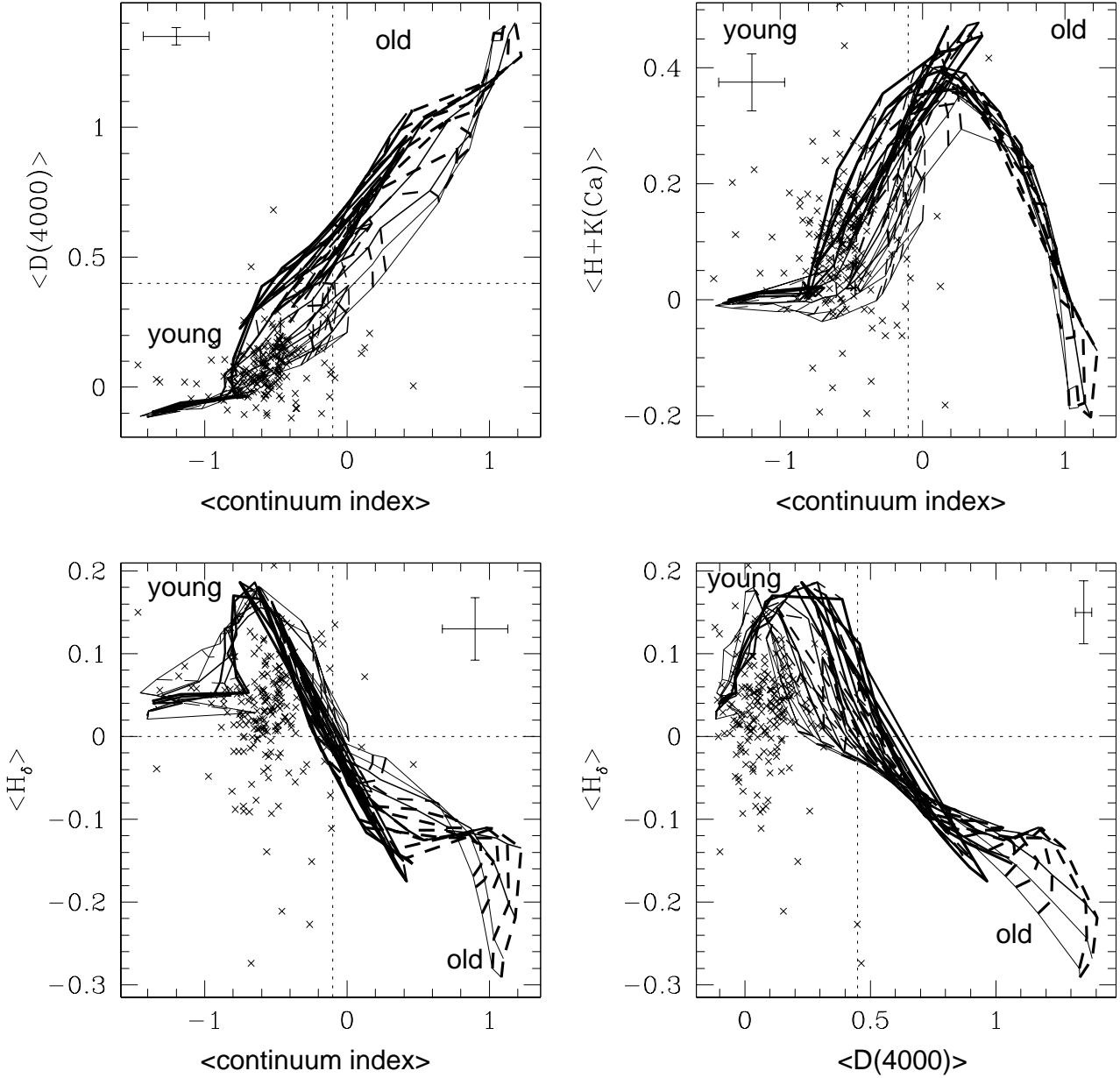}
    \caption{Distribution of the sample spectra (symbols) in different
        two-index planes compared with SSPs (lines). The solid lines
        represent the evolutionary paths of SSPs of different metallicities
        between ${\rm [Fe/H]}=-2.25$ and ${\rm [Fe/H]}=+0.35$, and the
        dashed lines represent the locus of SSPs of constant ages ranging
        from 0 to 20 Gyr. The dotted lines show the values of the indices
        at which SSPs of average metallicity (${\rm [Fe/H]}\sim-1$)
        reach an age of around 2 Gyr, to demarcate the ranges of
        young/intermediate populations and old populations.
        Young populations tend to be in the left part of the diagrams, old
        ones in the right part. The typical errors in the indices are shown
        as error bars in the corners of the different panels.}
    \label{featin_featin}
   \end{figure*}

\begin{figure*}
    \includegraphics[width=\textwidth]{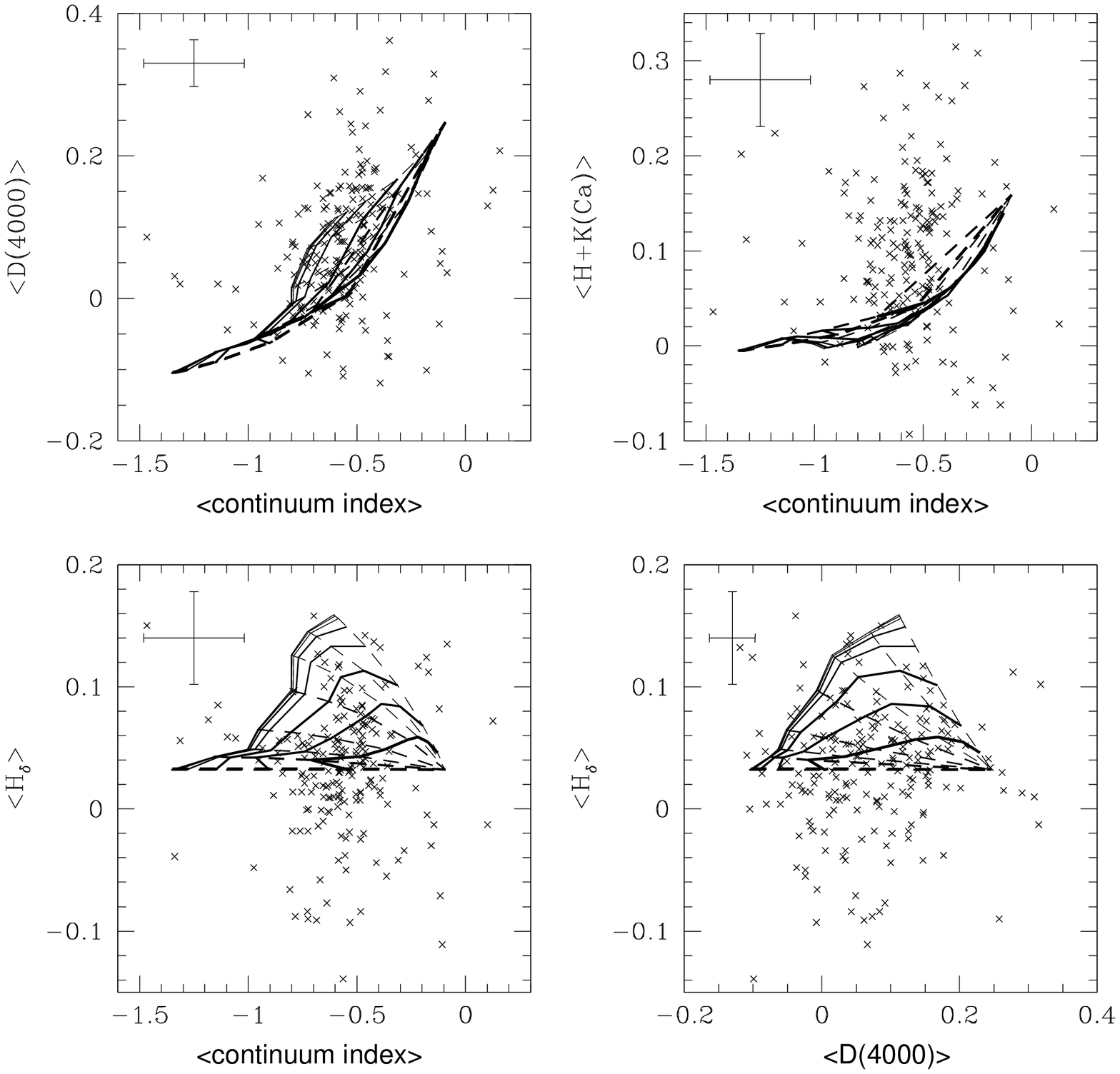}
    \caption{Distribution of the sample of spectra (symbols) in different
         two-index planes compared with composite stellar populations
         (lines). In relation to Fig.~\ref{featin_featin}, the range of
         the indices has been centered on the region where most of the
         observed data lie. The solid lines represent the compositions of
         an old population (age 5 Gyr, ${\rm [Fe/H]}=-1.5$) and a young
         or intermediate one (${\rm [Fe/H]}=-1$) with different mass ratios
         ranging from $M_{y+i}$:$M_{o} = 0:1$ (only an old population) to
         1:0 (only a young/intermediate population). The mass of the young
         population decreases with increasing line thickness, and the age
         of the young population varies along the lines. The dashed lines are
         lines of constant age of the young/intermediate population varying
         from 0 to 500 Myr (increasing line thickness means increasing age),
         whereas the mass ratio between the two populations varies along the
         lines. They converge towards the point where the mass of the
         young/intermediate population is zero. The typical errors in the
         indices are shown as error bars in the corners of the different
         panels.}
    \label{featin_featinb}
   \end{figure*}

   The average signal-to-noise of the spectra amounts to 12.8 at the
   level of the continuum around 5500 \AA\ \citep[for the signal-to-noise
   values of individual spectra, see][]{kehrig}.
   Using the relation that the error in the average flux in a passband
   amounts to the error per pixel divided by the square root of the
   number of pixels, we can calculate the typical observational
   errors of the narrow-band indices, whereas the typical error of
   the continuum index is dominated by uncertainties in the absolute
   flux calibration. Following \citet{cuisinier_96} and \citet{kong},
   we adopt a value of 10\% for these uncertainties.
   We list the typical observational error of each index in
   Table~\ref{indexranges}. They are also shown as error bars
   in Figs.~\ref{featin_featin} and \ref{featin_featinb}. \\
   Fig.~\ref{featin_featin} shows examples of how the spectra are
   distributed in two-index planes compared to the same indices
   calculated for simple stellar populations (SSPs) of a wide range of
   ages (dashed lines) and metallicities (solid lines) from the
   ``BC99'' (``Bruzual \& Charlot 99'') SSP library which will be
   described later on in this section.
   It can be seen that the observed indices (crosses) of the spectra
   lie in the region where the SSPs of young and intermediate age can
   be found (denoted by ``young'' in Fig.~\ref{featin_featin}), which
   shows that young and intermediate age populations dominate the
   spectra (but not necessarily the stellar mass of the galaxies). In
   the panels showing the continuum index vs. H$_{\delta}$ plane
   (lower left), and the panel showing the 4000 \AA\ break
   vs. H$_{\delta}$ plane (lower right), it is striking that the
   observed indices of many spectra lie outside the ranges occupied by
   the SSP models, eg. inside the loop in these ranges, another
   indication that one single population alone cannot reproduce the
   sample spectra adequately.

   An indication that this can be remedied by combining populations of
   different ages and metallicities can be seen in
   Fig.~\ref{featin_featinb}. It shows the same two-index planes
   as in Fig.~\ref{featin_featin} (with different scales), but here
   the empirical indices are compared with composite stellar
   populations. The composite populations are made up of an old
   population with ${\rm [Fe/H]}=-1.5$ and an age of 5 Gyr and, and a
   young or intermediate one with ${\rm [Fe/H]}=-1$ and different ages
   ranging from 1 to 500 Myr (different dashed lines, increasing line
   thickness means increasing age). The solid lines mean different
   mass ratios between the two populations ranging from
   $M_{y+i}$:$M_{o} = 0:1$ (only an old population) to 1:0 (only a
   young or intermediate population), increasing line thickness means
   decreasing $M_{y+i}$:$M_{o}$. The main problem of single stellar
   population models to explain the distribution of the observed data
   in the two-indice planes is that for index combinations containing
   ${\rm <H\delta>}$, the observed data fill a region where no models
   are present, within a loop created in the models at young ages (see
   Fig.~\ref{featin_featin}, lower panels).
   Combining young+intermediate and old populations in composite models
   makes it possible to resolve this issue, filling in the loop, and
   thus better reproducing the combination of indices of the sample
   spectra than SSP models are able to do
   (see Fig.~\ref{featin_featinb}, lower panels).
   For these reasons, we decided to explore the stellar composition of
   these galaxies by assuming
   them to be made up of two stellar populations, an old, metal-poor
   one, hereafter the ``old population'', and a younger one (that is,
   of young and/or intermediate age), which for simplicity will be
   referred to as the ``young+intermediate population''.

   This is a simplification, for the young+intermediate population
   will always contain the very young ionizing massive stellar
   population (age~${\rm < 10 Myr}$, possibly mixed up with an
   intermediate age population, which may consist of several
   subpopulations (that we are unable to discern)). The old
   population may also represent a combination of several, but old,
   populations.

   The star forming history of HII galaxies may be far more complex than
   what can be realistically modeled in a simple manner from their
   spectra. Hypotheses for  star formation histories are various,
   as briefly mentioned in the introduction; a bursting mode followed by
   quiescent phases, continuous star formation but sustained at a low
   intensity extending far back in past, or a unique single event.
   Many of these scenarios, however, are indiscernible from the point
   of view of a spectral analysis using only integrated spectra,
   as was also noticed by \citet{lilly}.
   We thus have to limit ourselves to an arbitrary, simpler choice of
   populations whose properties we believe we are able to
   distinguish.

   Within this choice, it is not possible to determine more than only a
   few parameters, and the results found using only two populations
   already furnish interesting results about the formation histories of
   the sample galaxies.
   For the five parameters $M_{y+i}$:$M_{o}$ (the mass relation between the
   young+intermediate - and the old population), $age_{y+i}$/$age_{o}$
   (the ages of each population), and ${\rm [Fe/H]}_{y+i}$/${\rm [Fe/H]}_{o}$
   (the respective metallicities), we took into consideration the possible
   values given in Table~\ref{parameters}.
   \begin{table}
   \begin{center}
      \caption{Possible values of the population parameters.}
      \label{parameters}
         \begin{tabular}{ll}
            \hline
 parameter & possible values \\
            \hline
 $M_{y+i}$:$M_{o}$    & 0:1, 1:100, 1:30, 1:10, 1:3, 1:1, 3:1, 10:1, 1:0 \\
 $age_{y+i}$          & 1, 2, 5, 10, 20, 50, 100, 200, 500 Myr \\
 ${\rm [Fe/H]}_{y+i}$ & taken from \citet{cuisinier} or -1 \\
 $age_{o}$          & fixed at 5 Gyr \\
 ${\rm [Fe/H]}_{o}$ & fixed at -1.5 \\
            \hline
         \end{tabular}
   \end{center}
   \end{table}
   Note that only two of these parameters, $M_{y+i}$:$M_{o}$ and
   $age_{y+i}$, are free (thus to be determined in this work), and the
   discussion about the solutions will mainly consist of interpreting
   the values found for these two parameters. The metallicity of the
   young population is given by the ${\rm [O/H]}$ values of the gas as
   determined from the emission lines by \citet{cuisinier} in a
   parallel article to this one, which deals with the emission part of
   the spectra, as the young population should be more or less coeval
   with the gas, whose metallicity is the metallicity attained now by
   these galaxies \citep{cuisinier,pagel}. In those cases where ${\rm
   [O/H]}$ of the gas could not be determined, we fixed ${\rm
   [Fe/H]}_{y+i}$ at -1, which is close to the average value of the
   other spectra. The two parameters concerning the old population,
   $age_{o}$ and ${\rm [Fe/H]}_{o}$, are also fixed at reasonable
   values for such a population (5 Gyr, resp. $-1.5$~dex). However,
   the analysis is not very sensitive to the exact values of these two
   parameters, since spectral features do not vary much in this
   parameter range. Note also that this approach allows
   solutions consisting of only an old - or only a young+intermediate
   population ($M_{y+i}$:$M_{o}=0$:1 or 1:0).

   The best fits were performed using two different libraries of SSPs
   in order to determine the reliability of the results. Another
   by-product of this study is to test if the new synthetic stellar
   spectral library BaSeL 3.1, which was calibrated using globular
   cluster photometrical data  to improve the reproduction of
   the continuum contribution of spectra of low metallicity stars
   (${\rm [Fe/H]} \stackrel{_<}{_\sim} -1$), can be used for young or
   intermediate stars and populations, thus for objects outside the
   parameter range of the objects used for its calibration. The first
   SSP library (hereafter the ``BC99'' library)  was
   produced using the Bruzual and Charlot 2000 Galaxy Isochrone
   Spectral Synthesis Evolution Library (GISSEL) code
   \citep{charlot_91, bruzual_93, bruzual_00}, implementing the Padova
   2000 isochrones \citep{girardi_00} combined with the BaSeL 3.1
   ``Padova 2000'' stellar library. We used a Salpeter initial mass
   function (IMF) from 1 to 100 $M_{\odot}$, to be able to make direct
   comparisons with the second SSP library, from now on called the
   ``Starburst'' library, which uses the same IMF. The ``Starburst''
   library consists of spectra from the STARBURST99 data package
   \citep{starburst} using the option of including nebular continuum
   emission (Fig. 1 on the STARBURST99 web site). It implements the
   predecessor of the BaSeL 3.1 library, the widely used BaSeL 2.2
   library \citep{lejeune_97, lejeune_98}, and for stars with strong
   mass loss the \citet{schmutz} extended model atmospheres, combined
   with the Geneva isochrones
   \citep{meynet,schaller,schaerera,schaererb,charbonnel}. Apart from
   the implementation of different isochrones and stellar libraries,
   one of the main differences between the ``BC99'' and the
   ``Starburst'' libraries lies in the inclusion of the nebular
   continuum emission under the hypothesis of optical thickness of the
   nebulae, e. g. that all photons with wavelengths below 912 \AA\ are
   absorbed. This affects the spectra of young populations (up to 10
   Myr) shortward of 4500 \AA . For the old population, the ``BC99''
   spectrum was used, since the Starburst99 data package only contains
   spectra up to 900 Myr. 

   The best fits were performed by minimising the quantity
   \begin{equation}
    \chi^{2}= \sum_{indices} \left(
    \frac{<i>_{emp}-<i>_{synth}}{\sigma_{<i>}} \right)^{2}
   \label{chidois}
   \end{equation}
   where $<i>_{emp}$ is the value of the
   $i$th index of the empirical spectrum, $<i>_{synth}$ is the value of
   the same index of the synthetic spectrum, produced by adding the
   synthetic spectra of the young+intermediate and the old
   population in the relation $M_{y+i}$:$M_{o}$,
   and $\sigma_{<i>}$ is the observational error in the $i$th index
   of the spectrum.
   As the Fe4531 and Mg$_b$ indices proved too affected by noise for our
   purpose, we decided not to adopt them in the fits (although the lines
   are clearly visible in many of the spectra, which is why they were
   originally included in the set). The fact that ${\rm [Mg/Fe]}$ is
   probably non-solar in these galaxies may cause additional complications
   as well if we use these indices.
   However, we did include them in the analysis performed in
   Sects.~\ref{featin_param} to \ref{BC99vsStarburst}.

   \begin{figure*}
    \includegraphics[width=\textwidth]{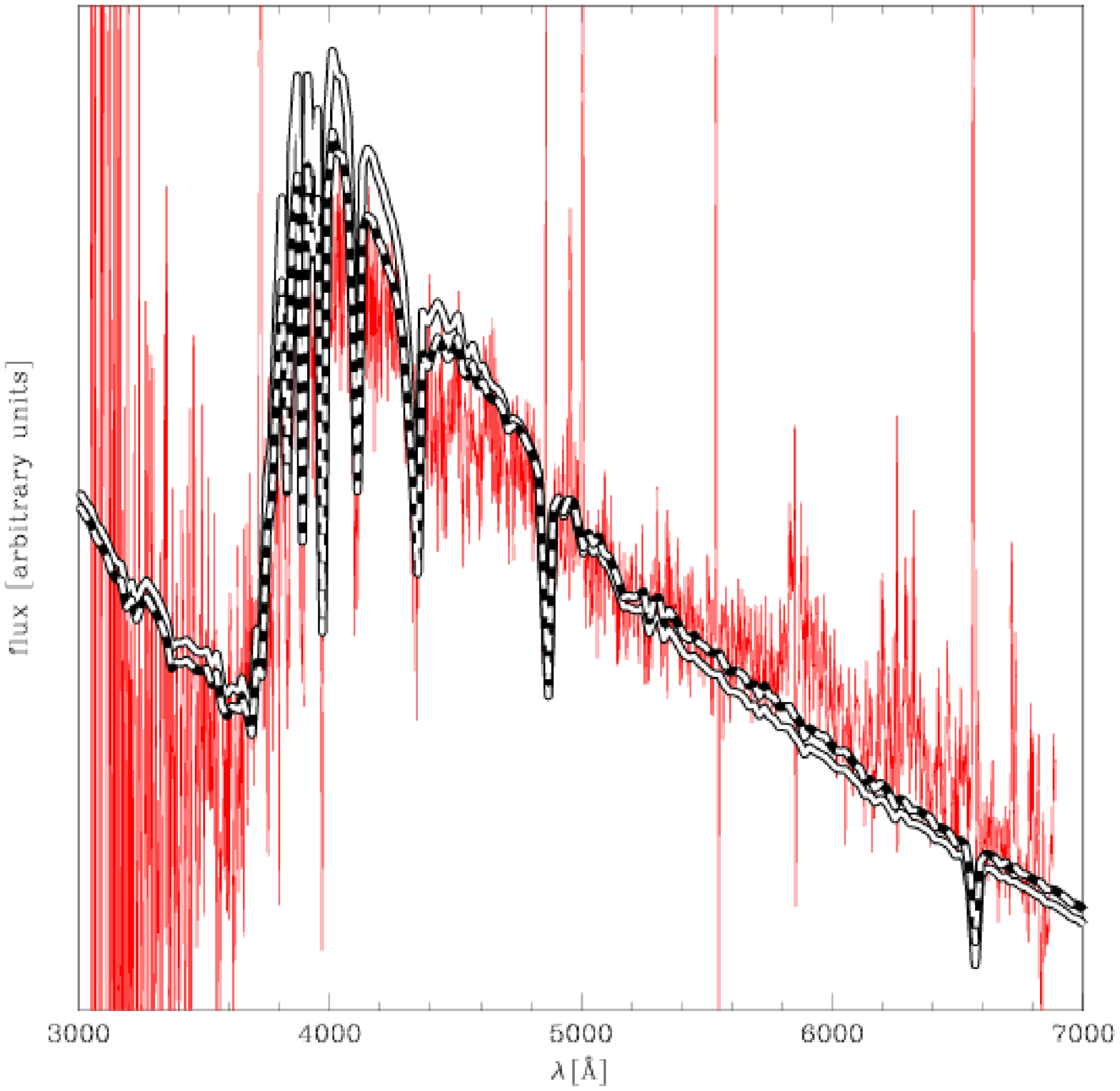}
    \caption{Example of a ``best fit''. The thin line represents
       the empirical spectrum (UM 137(W), taken on August, the 18th,
       1998), the thick solid white line shows the best
       fit spectrum obtained using the ``BC99'' library, and the thick
       dashed white line (which sometimes hides the solid line) shows the
       best fit using ``Starburst''.}
    \label{rejectedaccepted}
   \end{figure*}
   In the end, a selection of acceptable solutions was made. All spectra
   which had a too bad signal-to-noise or a strange shape,
   indicating calibration problems in the data reduction,
   were eliminated from the sample. This selection was done by eye.
   Fig.~\ref{rejectedaccepted} shows an example of an accepted fit.

   \begin{figure}
    \includegraphics[width=\columnwidth]{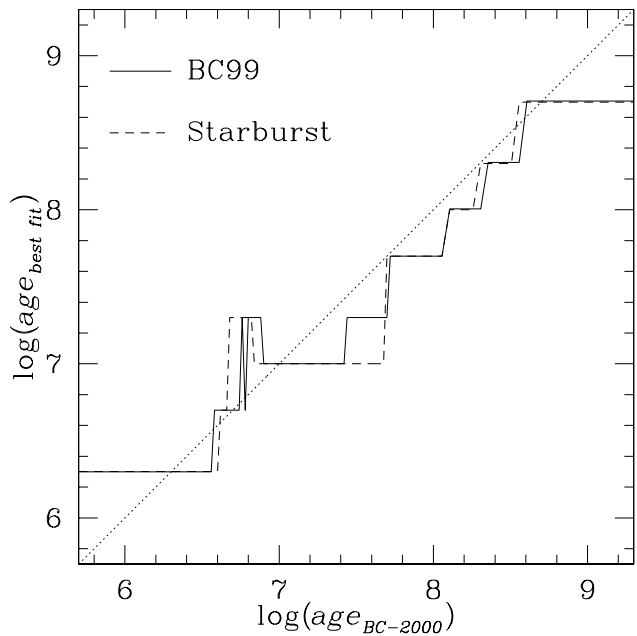}
    \caption{Ages of the best fits performed on ``BC-2000''
       synthetic SSPs as a function of their ages.
       The solid line shows the best fits using the
       ``BC99'' library, and the dashed line shows the best fits using
       ``Starburst''. The dotted line represents identity.}
    \label{agefits}
   \end{figure}
   A major concern in our method stems from the fact that the
   spectra of our sample have a higher resolution than the spectra in
   the ``BC99'' and ``Starburst'' libraries. Whereas the sample
   spectra have a resolving power of $R \sim 1000$, the synthetic
   spectra have only around $R \sim 250$ in the optical. However,
   degrading our spectra to this resolution would significantly
   reduce the amount of information they hold, particularly in the
   H$_{\delta}$ index, for which the contamination of the
   H$_{\delta}$ emission line would increase dramatically. As this
   line is in general much stronger than the absorption line which
   the index is trying to quantify, it would dominate the index after
   a degradation. Correcting for the contribution of the emission
   line using an estimate of the H$_{\delta}$ flux based on the
   fluxes of other Balmer lines would still leave us with an error as
   large as the error of this estimate, which is much larger than the
   variation of the index itself. For this reason, we decided to
   study the effect of resolution on the fits rather than to degrade
   our sample spectra. This study was done by performing the best
   fits on synthetic high-resolution spectra of known ages. The
   synthetic spectra, in the following called the ``BC-2000''
   synthetic SSP spectra, were again produced with the GISSEL
   software using the same input parameters (IMF, stellar
   evolutionary tracks) as for the ``BC99'' library but this time
   implementing the BC-2000 (Bruzual \& Charlot 2000) stellar
   library. This is a library of empirical stellar spectra, mainly
   from a catalog assembled by \citet{pickles}, with higher
   resolution than the theoretical libraries ($R \sim 500$), but
   containing only solar metallicity stars. For a more detailed
   description of the BC-2000 stellar library, see
   \citet{bruzual_00,bruzual_03}. We then determined the ages of
   these spectra in the same way as we determined the population
   properties of the HII galaxy spectra: by minimising the $\chi^{2}$
   defined in Eq. (\ref{chidois})
   (using a signal-to-noise value of 12.8 for the estimate of the
   errors $\sigma_{<i>}$ of the BC-2000 spectra, which corresponds
   to the average signal-to-noise of the HII galaxy spectra)
   for ``BC99'' and ``Starburst'' SSPs of solar metallicity, and
   considering ages of 1, 2, 5, 10, 20, 50, 100, 200, and 500 Myr.
   In Fig.~\ref{agefits}, the best
   fitting ages obtained in this way are shown as a function of the
   age of the original ``BC-2000'' SSP ages. In spite of the
   different resolutions of the ``BC-2000'' (input) spectra and the
   spectra used to fit the ages (``BC99'' and ``Starburst''), the
   ages are well reproduced with a slight tendency to favour young or
   intermediate ages. We conclude from this that the higher
   resolution of the empirical spectra does not pose greater problems
   for the fitting procedure and for the accuracy of the parameters
   we wish to determine. For these reasons,
   we preferred not to degrade the spectra of our sample to the
   ``BC99'' and ``Starburst'' resolution.

\section{Accuracy of the method}
\label{results}

   To check how useful our indices are for deducing
   population mass ratios and ages from our spectra we performed
   various tests with the observed indices and derived parameters.

   After rejection of all spectra that presented strange behaviour of
   the fitted model spectra, solutions were found for 123 spectra from
   78 different galaxies.

   Performing the best fits while varying the indices by their typical
   observational errors (as listed in Table~\ref{indexranges}) showed that
   in the worst case (if the errors in all indices conspire), they can
   modify the solutions by around one step (a factor of $\sim$3) in
   $M_{y+i}$:$M_{o}$ or two steps (a factor of $\sim$5) in $age_{y+i}$.
   For most spectra, however, the errors will
   not conspire, as they are random and not correlated,
   so the statistical properties of the sample discussed
   in this section should not be too affected by the observational
   errors. The solutions are presented in Table~\ref{solutions}.

\subsection{Correlations between the empirical indices and the obtained parameters}
\label{featin_param}

   \begin{figure*}
    \includegraphics[width=\textwidth]{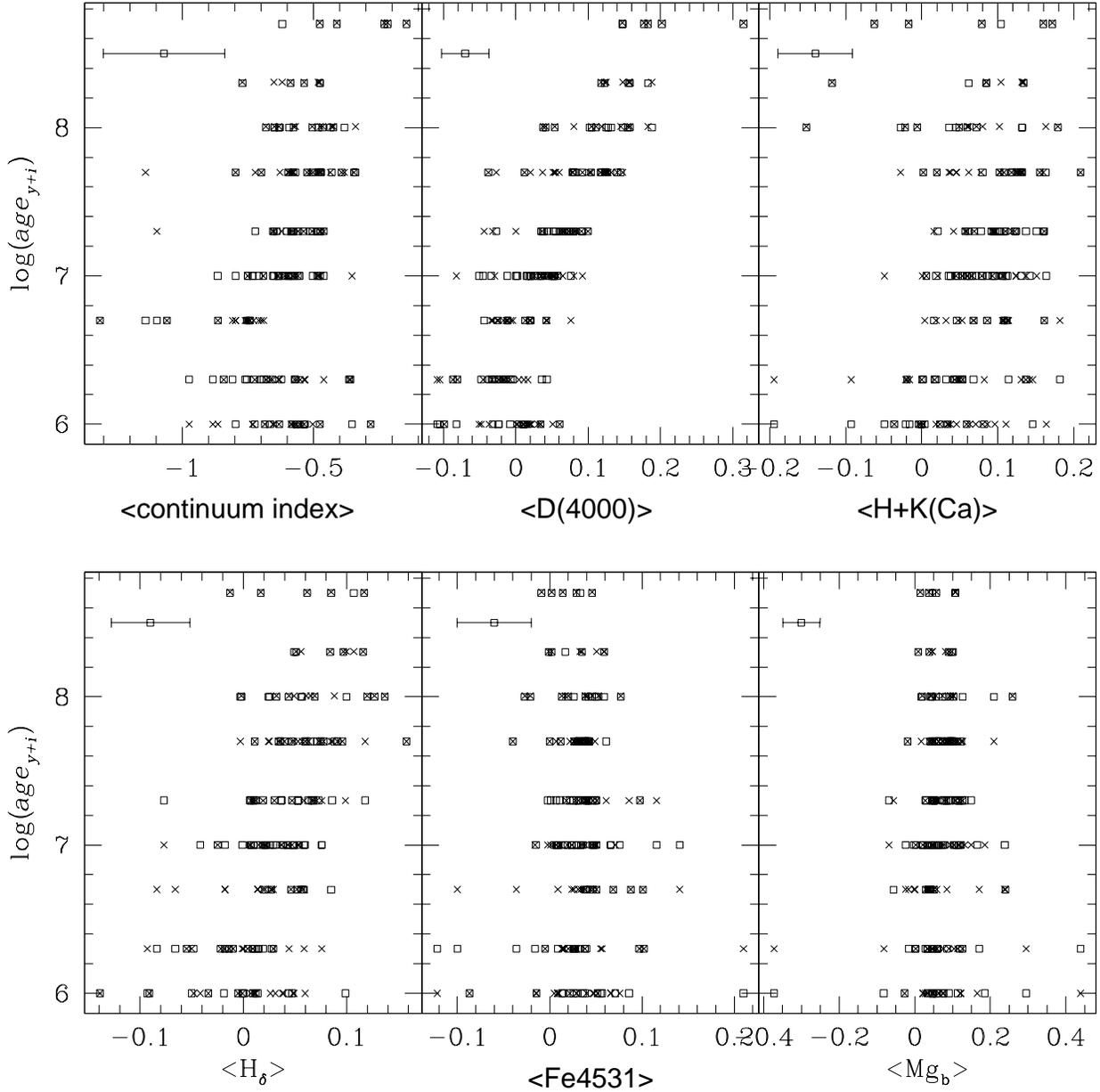}
    \caption{The ages of the young+intermediate population as a function of
       the indices of the spectra. The crosses represent the solutions
       found using ``BC99'', and the squares represent the ``Starburst''
       solutions.
       The typical errors in the indices are shown as error bars in the
       upper left corners of the different panels.}
    \label{featin_agey}
   \end{figure*}
   We will first investigate how these results depend upon the
   different indices of the galaxy spectra. $M_{y+i}$:$M_{o}$
   correlates (weakly) with the continuum index and with D(4000),
   but not with the other indices (not shown here). Fig.~\ref{featin_agey}
   shows the correlations of the indices with the derived age of the
   young+intermediate population. It correlates with D(4000) (upper
   middle) and (weakly) with the continuum index (upper left) and
   H$_{\delta}$ (lower left). For the continuum index,
   the trend reverses at $age_{y+i}=5$ Myr, where the
   nebular emission starts to influence the continuum (for very young
   ages it even dominates the spectra below 3600\AA).
   The other indices correlate only weakly (if at all) with $age_{y+i}$.

\subsection{Reproduction of the indices}
\label{featin_featout}

   \begin{figure*}
    \includegraphics[width=\textwidth]{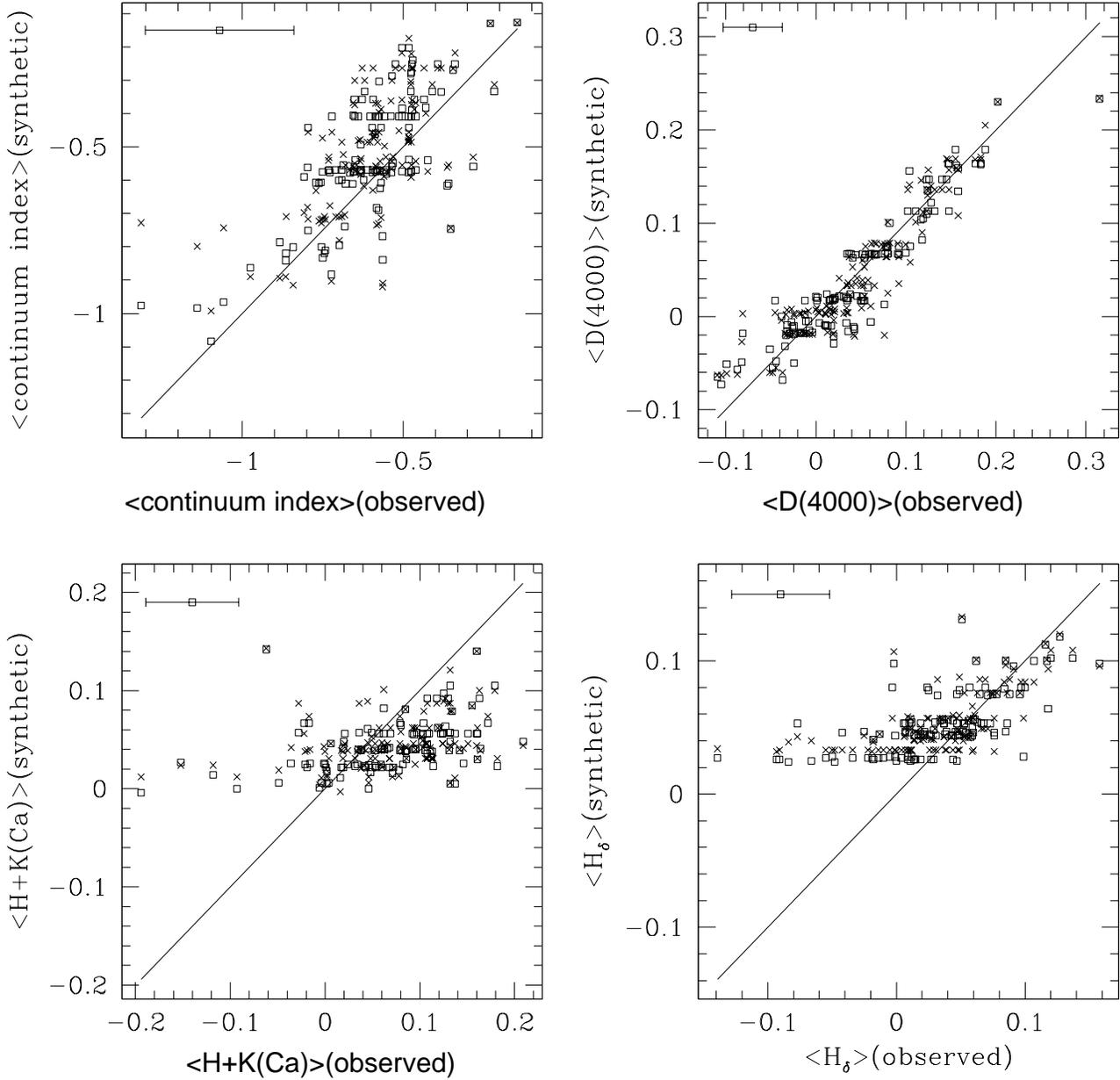}
    \caption{The indices of the empirical spectra vs. the same indices
       of the synthetic spectra (thus the reproduction of
       these indices). The crosses represent the solutions
       found using ``BC99'', and the squares represent the ``Starburst''
       solutions. The lines represent identity (perfect reproduction).
       The typical errors in the observed indices are shown as error bars
       in the upper left corners of the different panels.}
    \label{featinout}
   \end{figure*}
   The examination of how well the indices of the galaxy spectra are
   being reproduced in the fits yields the answer that could be
   expected from the typical observational errors of each index
   (Fig.~\ref{featinout}): The 4000 \AA\ break is well
   reproduced in the best fit spectra, and the continuum index
   a bit less well (but acceptably). The ${\rm H+K(Ca)}$
   and H$_{\delta}$ indices represent the worst cases of our
   adopted fit indices: our fit values present some correlation with
   the observed values, but with a systematic tendency to an
   underestimation of ${\rm H+K(Ca)}$ and an overestimation of
   H$_{\delta}$.
   For H$_{\delta}$, this is due to the fact that
   the models only predict values above $\sim 0.03$. This can also
   be seen in the lower panels of Fig.~\ref{featin_featinb}. 
   A possible explanation is that the underlying population in some
   of the galaxies could be older than 5 Gyr, which would lead to
   lower values of H$_{\delta}$.
   However, the trends in ${\rm H+K(Ca)}$ and H$_{\delta}$ are
   unlikely to influence the statistical properties of our overall
   results in a systematic manner, as they drive the parameter
   $M_{y+i}$:$M_{o}$ in opposite directions, as well as the
   age of the young+intermediate population.

\subsection{Comparison of the solutions obtained with the two different libraries}
\label{BC99vsStarburst}

   \begin{figure}
    \includegraphics[width=\columnwidth]{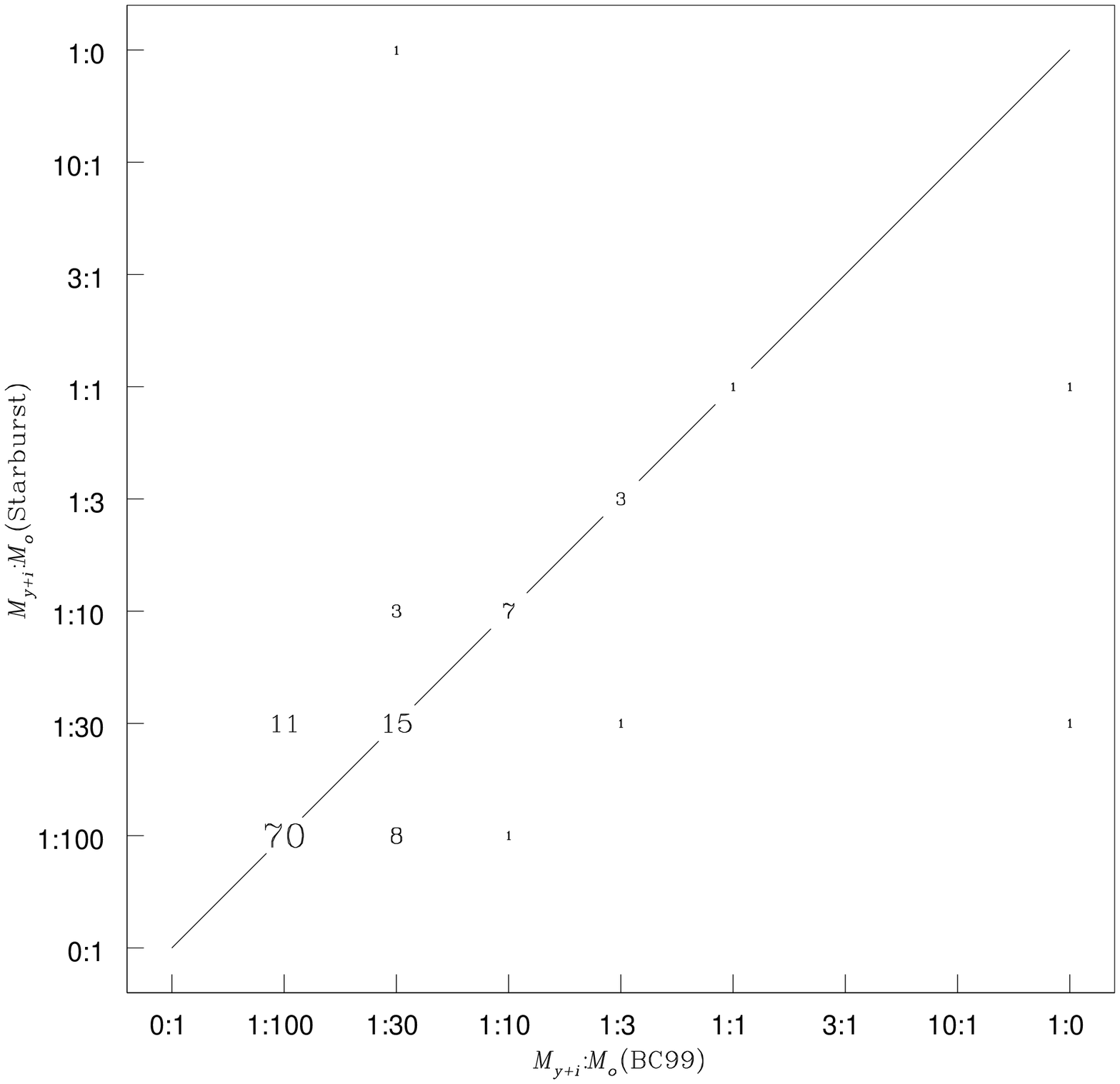}
    \caption{The number of cases, in which a certain value ($x$) of
       $M_{y+i}$:$M_{o}$ was found using ``BC99'' and another ($y$)
       value was found using ``Starburst''.}
    \label{Mybcvssb}
   \end{figure}
   \begin{figure}
    \includegraphics[width=\columnwidth]{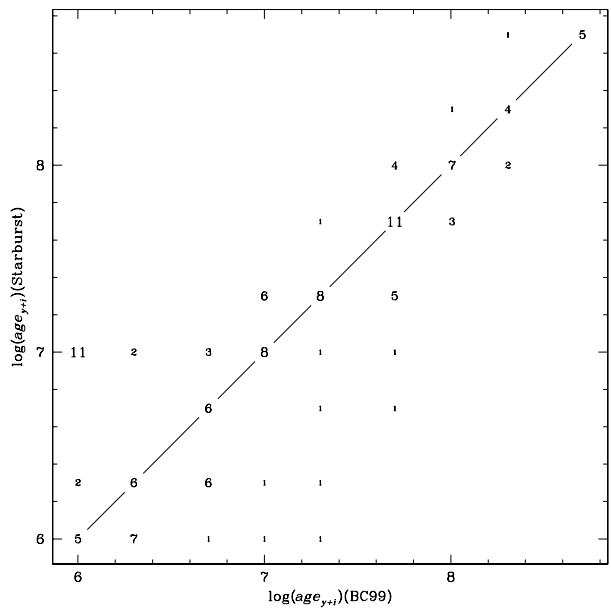}
    \caption{The number of cases, in which a certain ($x$) value of
       $age_{y+i}$ was found using ``BC99'' and another ($y$)
       value was found using ``Starburst''.}
    \label{ageybcvssb}
   \end{figure}
\begin{figure*}
    \includegraphics[width=\textwidth]{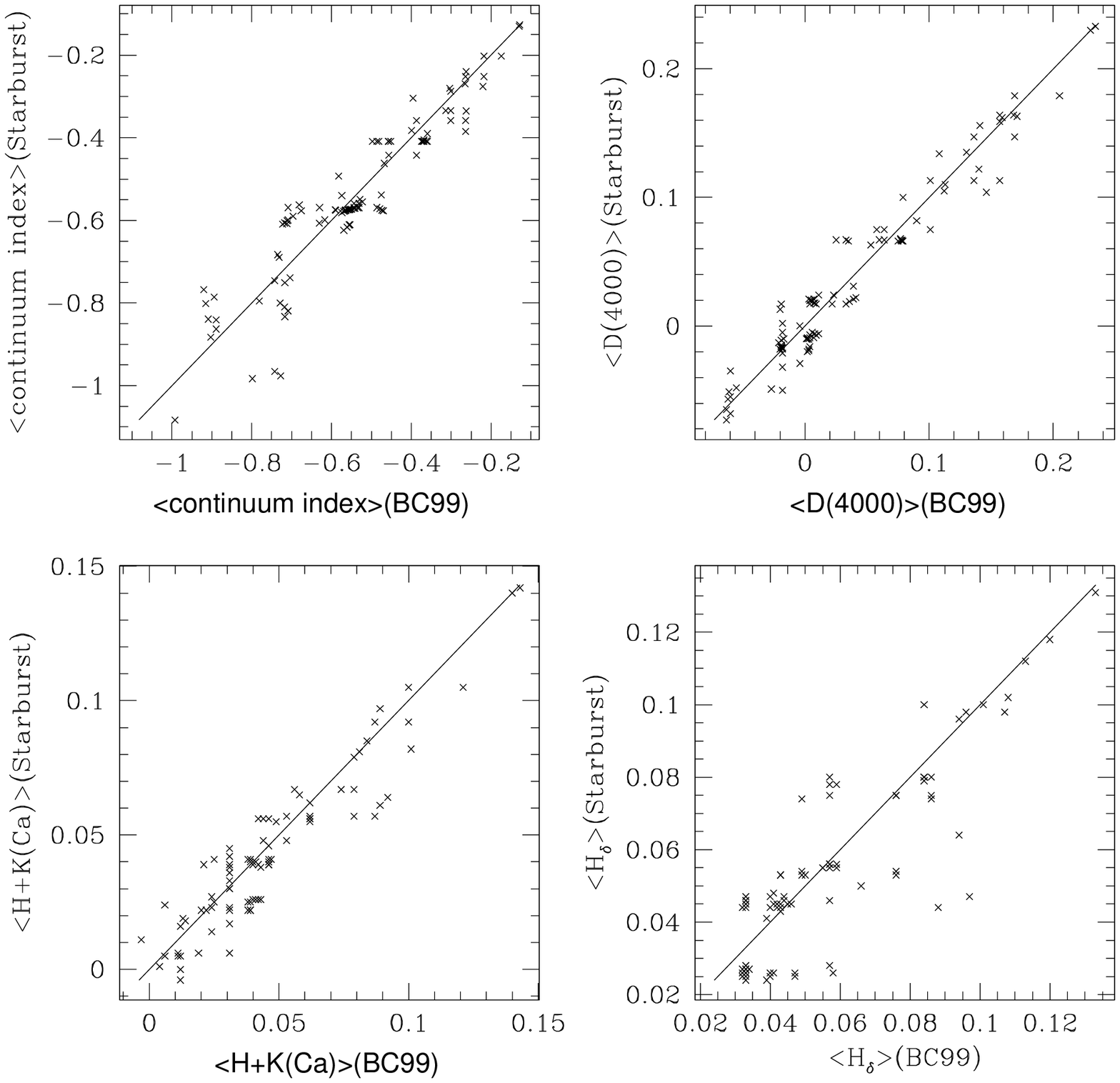}
    \caption{Comparison of the indices of the solutions obtained with
       ``BC99'' and ``Starburst'' (crosses).
       The lines represent identity.}
    \label{featbcvssb}
   \end{figure*}

   To determine to what amount these results depend on the SSP library
   (and thus on the stellar library) that was used, it is necessary to
   compare the results obtained with the two different libraries for the
   same spectra. 
   Fig.~\ref{Mybcvssb} shows the agreement between the best fits made
   with the two libraries for the $M_{y+i}$:$M_{o}$ ratio, and
   Fig.~\ref{ageybcvssb} shows the same for the $age_{y+i}$.
   In most cases, the two libraries yield the same value for
   $M_{y+i}$:$M_{o}$, and this parameter differs significantly between
   the two libraries only in three cases.
   In these three cases, either the ``BC99'' - or the ``Starburst''
   solution predicts the existence of only a young+intermediate
   population ($M_{y+i}$:$M_{o}=1$:0), whereas the other library
   predicts a significant old population ($M_{y+i}$:$M_{o}=1$:1 or 1:30).
   A closer look at these spectra does not reveal why the fits with the
   two libraries yield such different results. In spite of the completely
   different population parameters, the best fitting composite model
   spectra of the ``BC99'' and ``Starburst'' solutions look very similar.
   In almost half of the cases however, the two libraries yield
   identical results, and in most of the rest the solutions are similar
   enough.
   The two libraries also do not show any systematic differences in the
   indices calculated from the best fit spectra
   (see Fig.~\ref{featbcvssb}).  \\
   \begin{figure}
    \includegraphics[width=\columnwidth]{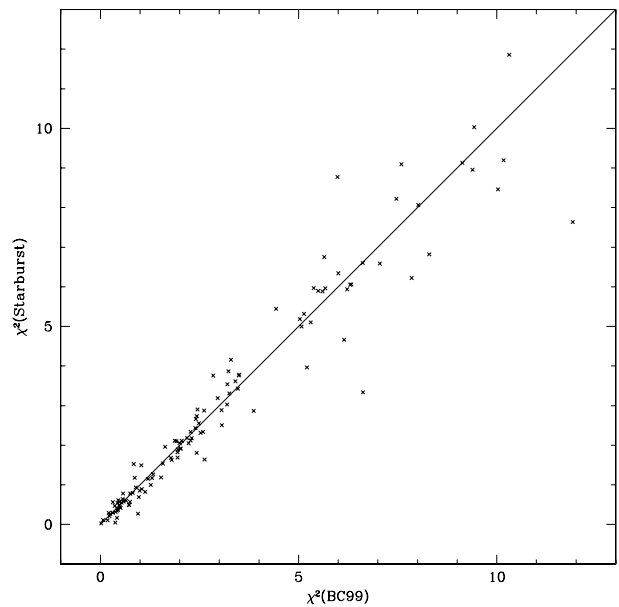}
    \caption{Comparison of the $\chi^{2}$ of the solutions obtained with
       ``BC99'' and ``Starburst'' (crosses).
       The line represents identity.}
    \label{CHIsqubcvssb}
   \end{figure}
   The similarities between the solutions found with the two libraries
   are also confirmed when comparing the $\chi^{2}$
   (Fig.~\ref{CHIsqubcvssb}). \\
   Thus, for the vast majority of the spectra,
   the solutions found with both libraries show good
   agreement in all three: the population parameters found, the
   reproduction of the indices, and $\chi^{2}$, which gives us
   confidence in the solutions and in our general conclusions.
   Given all these similarities, it is not possible to conclude from
   this study alone if the BaSeL 3.1 stellar library does a better job
   at reproducing spectra than its predecessor. It is encouraging that
   the two SSP libraries yield results of comparable quality, even though
   the ``Starburst'' library has additional features implemented (nebular
   continuum emission, inclusion of the Schmutz model atmospheres for stars
   with strong mass loss) aimed at a better reproduction of the spectra of
   young+intermediate populations.
   Further studies, aimed more directly at the properties for which the
   BaSeL 3.1 library was designed (a better reproduction of the continuum
   contribution of low metallicity stellar spectra)
   will be necessary to clarify this point.

\section{Results and discussion}
\label{distribution}
   \begin{figure}
    \includegraphics[width=\columnwidth]{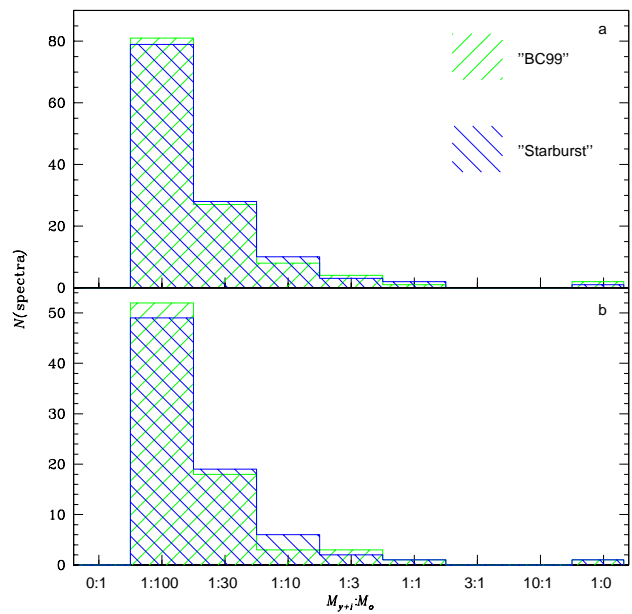}
    \caption{Histogram of the best fitting $M_{y+i}$:$M_{o}$ for the
       two different libraries. {\em upper panel}: all apertures in all
       galaxies. {\em lower panel}: one aperture (the lowest
       $M_{y+i}$:$M_{o}$ value) for each galaxy.}
    \label{Myhisto}
   \end{figure}
   \begin{figure}
    \includegraphics[width=\columnwidth]{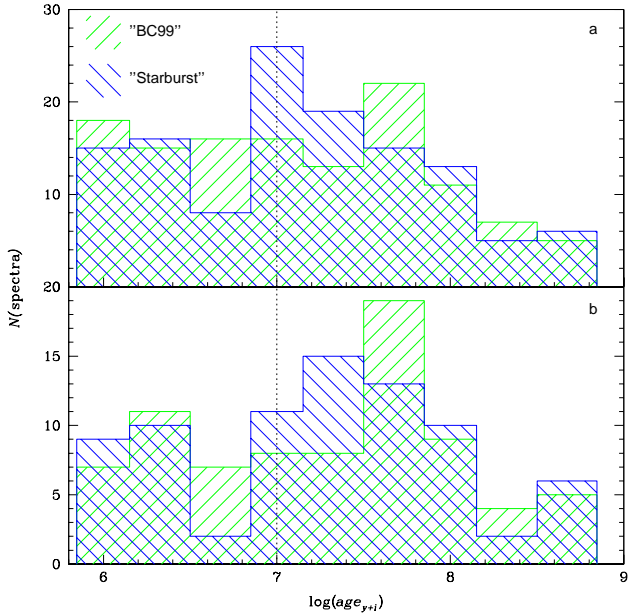}
    \caption{Histogram of the best fitting $\log(age_{y+i})$ using the
       two different libraries. {\em upper panel}: all apertures in all
       galaxies. {\em lower panel}: one aperture (the highest
       young+intermediate age) for each galaxy.
       The dotted line corresponds to the maximum duration of an SSP
       burst, which means that for ages younger than $10^{7}$ yrs, the
       young+intermediate population found could consist of only
       the young component.}
    \label{ageyhisto}
   \end{figure}

   The composition of HII galaxies may be far more complex than just the two
   young+intermediate and old populations we are using in our modeling
   --- as we stated earlier, they can have a far more complicated star
   forming history, not only in time but also in its spatial
   distribution. Our modeling represents our lack of ability to segregate
   more than only a few stellar populations with very distinct ages. It
   is in particular very hard to differentiate old populations because
   their spectrophotometric properties are so similar \citep{lilly};
   that is the reason why we grouped them all in a generic old
   population, but we have to keep in mind that we have very little
   insight in the constitution of this generic old population.

   Nevertheless, some star forming history scenarios can clearly be
   ruled out from the point of view of our modeling, such as extreme
   limiting scenarios presenting one single population, either only old
   or only young.

   In this section, we will discuss various possible star forming
   histories in the light of our modeling.

\subsection {Mass ratios of the young+intermediate and old populations}

   We were able to derive mass ratios between the young+intermediate
   and the old populations. Fig.~\ref{Myhisto} shows how these mass
   ratios are distributed. The upper panel is a histogram of the best
   fitting $M_{y+i}$:$M_{o}$ of all apertures for all galaxies for which
   solutions were found, and the lower panel is a histogram made up of
   one $M_{y+i}$:$M_{o}$ value (the lowest one found) per galaxy, in
   order to emphasise the old stellar component.

   Our study indicates that in terms of mass, the old population is
   overwhelmingly dominant, by a factor of $\simeq 100$ in most
   of the apertures in all galaxies (the peaks at
   $M_{y+i}$:$M_{o}=1:100$ of Fig.~\ref{Myhisto}).

   For the one galaxy where the integrated spectrum yielded
   only a young+intermediate population (DDO155) we were able to
   obtain evidence for an old population in other apertures.
   Only in two cases that were observed in one aperture only (Tol1223-359
   and UM 323) did the best fits yield no old population
   (the sole points at $M_{y+i}$:$M_{o}=1:0$ of panel b of
   Fig.~\ref{Myhisto}), albeit only in the fit using one of the two
   libraries.
   In these cases, the fit with the other library did present a
   (dominating) old population (see also Sect.~\ref{BC99vsStarburst}),
   so even in these dubious cases we have some evidence for an old
   population.

   It is however difficult to make any further statement on the nature of
   the old population, except that it should be older than
   $\simeq$~1~Gyr, because spectral features are so similar after this
   age, when compared to younger populations --- which actually dominate
   the spectra of these galaxies. Within this age limitation, the old
   population could be anything: accumulation of former bursts, or a low
   intensity, but long lasting continuous star formation episode.

   Whatever the star forming history of the old population may be, the
   important point is that the sum of the early events of star formation
   turned much more gas into stars than the present one.

\subsection {Age of the young+intermediate population}

   The only other free parameter of our study is the age of the
   young+intermediate population. Fig.~\ref{ageyhisto} shows histograms
   of the best fitting $age_{y+i}$ of all spectra for which
   solutions were found (upper panel), and of the oldest young+intermediate
   population found in each galaxy (lower panel).

   The young+intermediate population is a mixture of a certain young
   bursting population that is able to ionize the interstellar medium
   (age $<$~10~Myr) and - maybe - an intermediate age population (whose
   age could be anything between 10~Myr and 500~Myr).

   It is noteworthy that a significant fraction of our galaxies ($\simeq$
   60\%) present young+intermediate ages greater than 10~Myr (lower panel
   of Fig.~\ref{ageyhisto}), indicating in these cases the very
   existence of an intermediate age population in addition to the
   ionizing massive stellar component which necessarily is younger than
   10~Myr. Some individual differences exist between the fits with the
   ``Starburst'' and ``BC99'' libraries, especially in the age
   range from 5 to 50 Myr.
   However, in general the results  are consistent, as could already
   be seen at the end of the previous section.

   As we stated earlier, we do not disentangle possible multiple events
   in the intermediate age population but only indicate its presence and
   its average age. Therefore, it does not appear in our modeling by
   itself, but rather in a weighted mean together with the young
   population --- with weights that are difficult to determine.
   Despite these drawbacks, some points are, however, worth being stated.
   First, the mean age distribution of the young+intermediate population,
   although quite broad, is peaked around 10/20~Myr (``BC99'') and
   50~Myr (``Starburst''), indicating
   that the mean age of the intermediate population should be at least of
   the same order.

   To summarise, in 60\% of the galaxies of our sample, we see an
   intermediate population, whose mean age should be of the order of at
   least $\simeq$~20/50~Myr and more likely $\geq$~50~Myr.

\section{Summary and conclusions}
\label{conclusions}

   In this work, we analyse the stellar content of a sample of HII
   galaxies, of which we have intermediate resolution ($R \sim 1000$)
   and high enough signal-to-noise spectra with a wavelength range of
   3700 to 7500 \AA, using the continua and certain absorption features
   of the spectra.
   To identify stellar populations in these galaxies and to
   determine their properties (masses, ages, metallicities), we defined
   six spectral indices (a continuum index and five indices inspired by
   Lick indices as defined by \citet{worthey}, and \citet{worthey_2}, but
   optimised for the sample spectra).
   In a best fit procedure, we then determined for each HII galaxy
   spectrum the combination of two synthetic SSP spectra 
   (one of an old stellar population, and one of a young/intermediate age)
   that best reproduces the indices of the empirical spectrum.

   Although there are certainly more than just two populations in our
   galaxies, we justify our approach by the fact that we want to use
   a minimum set of free parameters to describe their observed spectral
   features.\\

   The main results of the analysis are the following:\\

   (1) In {\em all} galaxies of our sample, 
   we detected an old, underlying population ($\geq 1$ Gyr), {\em
   generally dominating the stellar mass by a large factor}.
   Although we did not have any extreme metal-poor dwarf galaxies
   (O/H$<$7.5), which are the most suspected to possess only a young
   population, some compact HII galaxies were present in our sample,
   like Fairall~30. Albeit not the least evolved, these galaxies are
   the ones where the detection of an old population is most
   difficult, because the young population is also spatially dominant.

   (2) In most of our spectra ($\simeq 60$\%), we found evidence for
   an intermediate age population (e.g. $\geq 20$ Myr or more likely
   $\geq 50$ Myr). Because this population cannot be the very young
   ionizing population (with an age of ${\rm \leq 10}$ Myr) that we know to
   be present, this indicates the presence of at least two populations (a
   young and an intermediate one) connected to the present star forming
   event. \\

   Our analysis reveals that any realistic modeling of the stellar
   populations in HII galaxies and subsequent derivation of their
   physical and evolutionary properties must take into account the
   presence of at least these three average populations (a young, an
   intermediate and an old one).

\begin{acknowledgements}
   This work was supported by the Swiss National Science Foundation.
   Furthermore, we would like to thank the Funda\c{c}\~{a}o Carlos Chagas
   Filho de Amparo \`{a} Pesquisa do Estado do Rio de Janeiro (Faperj) for
   the help on the infrastructure for this project.
   We also thank the group ``Galaxies: Formation, Evolution, and Activity''
   of the programme PRONEX of the Conselho Nacional de Desenvolvimento
   Cientifico Tecnol\'{o}gico in Brazil for travel expense allowances.
   Finally we are grateful to an anonymous referee whose comments greatly
   helped us to improve the presentation of this paper. 
  
\end{acknowledgements}

\appendix

\section{Indices and population parameters of individual spectra}

\begin{table*}
      \caption{Indices of individual spectra.}
      \label{indices}
         \begin{tabular}{lrrrrrrr}\hline
            \hline
galaxy & type  & cont. i. & D(4000) & H+K(Ca) & H$_{\delta}$ & Fe4531 & Mg$_b$ \\\hline

Cam08-28A(W)  & re &  0.158 &  0.207 & -0.182 &  0.333 & -0.023 &  0.057 \\
Cam08-28A(Cent) & re & -0.648 &  0.054 &  0.072 &  0.120 & -0.027 &  0.093 \\
Cam08-28A(E)  & re & -0.691 &  0.050 &  0.079 &  0.043 &  0.034 &  0.116 \\
Cam08-28A    & in & -0.751 &  0.052 &  0.066 &  0.054 &  0.007 &  0.103 \\
Cam0357-3915  & in & -0.108 &  0.066 &  0.070 & -0.111 &  0.237 & -0.210 \\
Cam0840+1044  & in & -0.681 &  0.038 & -0.006 &  0.127 &  0.020 &  0.044 \\
Cam0840+1201  & in & -0.569 & -0.029 &  0.001 &  0.015 & -0.016 &  0.119 \\
Cam1543+0907  & in & -0.532 & -0.008 &  0.146 & -0.093 &  0.210 &  0.295 \\
CTS1006      & in & -0.655 &  0.140 &  0.066 &  0.045 &  0.037 &  0.058 \\
CTS1008      & in & -0.564 & -0.099 & -0.004 & -0.139 & -0.014 &  0.048 \\
CTS1011      & in & -0.572 & -0.033 &  0.050 & -0.022 &  0.097 &  0.126 \\
CTS1013      & in & -0.482 &  0.117 &  0.172 &  0.014 &  0.055 &  0.045 \\
CTS1016      & in & -0.602 &  0.039 &  0.151 &  0.053 &  0.038 &  0.030 \\
CTS1017      & in & -0.351 &  0.362 &  0.315 &  0.222 &  0.111 & -0.150 \\
CTS1018      & in & -0.639 &  0.092 &  0.108 & -0.077 &  0.007 &  0.149 \\
CTS1019      & in & -0.551 & -0.024 &  0.082 & -0.050 &  0.014 &  0.076 \\
CTS1020      & in & -0.516 &  0.682 &  0.055 &  0.369 & -0.008 & -0.025 \\
CTS1022      & in & -0.585 &  0.034 &  0.164 & -0.042 &  0.066 &  0.022 \\
CTS1028      & in & -0.842 & -0.087 &  0.137 &  0.029 & -0.005 &  0.003 \\
CTS1029      & in & -0.461 & -0.045 &  0.142 &  0.076 &  0.029 &  0.059 \\
CTS1033      & in & -0.474 &  0.183 &  0.062 &  0.049 &  0.017 &  0.101 \\
CTS1034      & in & -0.484 &  0.044 &  0.124 &  0.037 &  0.001 &  0.095 \\
CTS1035      & in & -0.392 & -0.119 & -0.196 &  0.132 &  0.038 & -0.029 \\
DDO060(NW)   & re & -0.535 &  0.158 &  0.134 &  0.097 &  0.059 &  0.098 \\
DDO060(SE)   & re & -0.476 &  0.177 &  0.172 &  0.085 &  0.002 &  0.108 \\
DDO060       & in & -1.338 &  0.031 &  0.202 & -0.039 & -0.183 &  0.149 \\
DDO070(W)    & re & -0.534 &  0.110 &  0.185 &  0.031 &  0.000 & -0.156 \\
DDO070(E)    & re & -0.558 &  0.080 &  0.137 &  0.006 &  0.011 & -0.068 \\
DDO070       & in & -0.430 &  0.143 &  0.262 &  0.074 &  0.025 & -0.238 \\
DDO075(NE)   & re & -0.821 &  0.058 &  0.049 &  0.057 &  0.193 &  0.263 \\
DDO075(CentNE)  & re & -0.788 &  0.033 &  0.137 &  0.097 &  0.125 &  0.303 \\
DDO075(CentSW)  & re & -0.864 &  0.042 &  0.161 &  0.046 &  0.088 &  0.240 \\
DDO155(NE)   & re & -0.726 &  0.126 &  0.175 & -0.018 &  0.026 & -0.015 \\
DDO155(SW)   & re & -0.797 &  0.076 &  0.113 & -0.018 &  0.034 & -0.024 \\
DDO155       & in & -1.097 & -0.044 &  0.016 &  0.059 &  0.039 & -0.056 \\
ESO289IG037  & in & -0.482 &  0.188 &  0.132 &  0.056 &  0.051 &  0.047 \\
ESO533G014   & in & -0.503 &  0.152 &  0.184 &  0.082 &  0.049 &  0.073 \\
Fairall30    & in & -0.585 &  0.091 &  0.110 &  0.010 &  0.031 &  0.078 \\
Haro24       & in & -0.721 &  0.060 &  0.062 &  0.053 &  0.031 &  0.053 \\
Haro25       & in & -0.524 &  0.123 &  0.108 &  0.036 &  0.032 &  0.063 \\
Haro30       & in & -0.345 &  0.124 &  0.155 &  0.060 &  0.030 &  0.095 \\
IC5154(N)    & re & -0.627 &  0.148 & -0.028 & -0.003 &  0.049 &  0.127 \\
IC5154(S)    & re & -0.453 &  0.193 &  0.140 &  0.069 &  0.058 &  0.057 \\
IC5154       & in & -0.491 &  0.174 &  0.096 &  0.064 &  0.057 &  0.082 \\
Marseille01  & in & -0.524 &  0.151 &  0.038 &  0.027 &  0.043 &  0.130 \\
Marseille68  & in & -0.392 &  0.126 &  0.118 &  0.011 &  0.043 &  0.104 \\
Marseille88  & in & -0.576 &  0.012 &  0.002 &  0.091 &  0.012 &  0.041 \\
MBG00463-0239  & in & -0.420 &  0.181 &  0.212 &  0.018 &  0.036 &  0.089 \\
MBG02411-1457  & in & -0.369 &  0.318 &  0.258 &  0.102 &  0.061 &  0.093 \\
MBG20533-4410  & in & -0.282 &  0.034 & -0.036 & -0.034 &  0.022 & -0.027 \\\hline

\end{tabular}
\end{table*}
\addtocounter{table}{-1}
   \begin{table*}
      \caption{Indices of individual spectra (continued).}
         \begin{tabular}{lrrrrrrr}\hline

galaxy & type & cont. i. & D(4000) & H+K(Ca) & H$_{\delta}$ & Fe4531 & Mg$_b$ \\\hline

MBG21567-1645(E)  & re & -0.527 &  0.245 &  0.195 &  0.034 &  0.088 &  0.129 \\
MBG21567-1645(Cent)  & re & -0.606 &  0.309 &  0.287 &  0.010 &  0.069 &  0.125 \\
MBG21567-1645(W)  & re & -0.580 &  0.262 &  0.251 &  0.030 &  0.071 &  0.130 \\
MBG21567-1645  & in & -0.504 &  0.155 &  0.179 &  0.044 &  0.077 &  0.103 \\
MBG22012-1550(E)  & re & -0.796 &  0.079 &  0.132 &  0.096 &  0.032 &  0.064 \\
MBG22012-1550(W)  & re & -0.742 &  0.127 &  0.153 &  0.042 &  0.074 &  0.041 \\
MBG22012-1550  & in & -0.589 &  0.126 &  0.148 &  0.055 &  0.059 &  0.039 \\
MBG23121-3807  & in & -0.490 &  0.209 &  0.181 &  0.052 &  0.050 &  0.080 \\
MCG0157017   & in & -0.474 &  0.035 &  0.124 &  0.053 &  0.049 &  0.073 \\
Mrk36        & in & -0.699 & -0.004 &  0.032 &  0.027 &  0.009 & -0.016 \\
Mrk710(SW)   & re & -1.183 & -0.300 &  0.224 &  0.073 & -0.026 & -0.442 \\
Mrk710(CentSW)  & re & -0.170 &  0.278 &  0.193 &  0.112 & -0.172 &  0.167 \\
Mrk710(CentNE)  & re & -0.594 &  0.054 &  0.038 &  0.049 &  0.007 &  0.111 \\
Mrk710(NE)   & re & -0.622 &  0.036 &  0.042 &  0.009 &  0.028 &  0.061 \\
Mrk710       & in & -0.611 &  0.035 &  0.053 &  0.021 &  0.036 &  0.051 \\
Mrk711       & in & -0.309 &  0.147 &  0.274 & -0.042 &  0.041 &  0.104 \\
Mrk1201      & in & -0.250 &  0.212 &  0.308 & -0.151 & -0.012 &  0.167 \\
Mrk1318      & in & -0.503 &  0.117 &  0.127 &  0.047 &  0.027 &  0.081 \\
NGC7323(E)   & re & -0.465 &  0.111 &  0.049 &  0.069 &  0.013 &  0.259 \\
NGC7323(W)   & re & -0.435 &  0.132 &  0.036 &  0.024 &  0.026 &  0.209 \\
NGC7323      & in & -0.641 &  0.129 & -0.001 &  0.018 &  0.071 &  0.032 \\
POX186       & in & -0.424 &  0.087 & -0.251 &  0.037 & -0.054 &  0.250 \\
Tol0104-388(NW)  & re & -0.563 & -0.109 & -0.093 &  0.044 &  0.057 &  0.090 \\
Tol0104-388(SE)  & re & -0.952 &  0.104 & -0.017 &  0.040 &  0.048 &  0.065 \\
Tol0117-414EW  & re & -0.575 &  0.037 &  0.160 &  0.086 &  0.022 &  0.115 \\
Tol0117-414NS(N)  & re & -0.731 &  0.051 &  0.111 & -0.001 &  0.009 &  0.120 \\
Tol0117-414NS(CentN)  & re & -0.585 &  0.101 &  0.194 &  0.056 &  0.047 &  0.063 \\
Tol0117-414NS(CentS)  & re & -0.484 &  0.065 &  0.021 &  0.037 & -0.002 &  0.132 \\
Tol0117-414NS(S)  & re & -0.772 &  0.018 &  0.273 &  0.060 &  0.118 & -0.004 \\
Tol0140-420  & in & -0.474 &  0.002 &  0.059 &  0.060 &  0.076 &  0.122 \\
Tol0226-390  & in & -0.630 &  0.011 & -0.021 &  0.009 &  0.013 &  0.114 \\
Tol0306-405  & in & -0.654 &  0.031 &  0.019 &  0.010 &  0.048 &  0.063 \\
Tol0341-407(E)  & re & -0.559 &  0.000 &  0.095 &  0.076 &  0.116 &  0.054 \\
Tol0341-407(W)  & re & -0.475 & -0.032 &  0.059 &  0.099 &  0.086 &  0.057 \\
Tol0341-407  & in & -0.179 & -0.101 & -0.044 &  0.124 &  0.070 &  0.080 \\
Tol0440-381  & in & -0.582 & -0.027 &  0.078 &  0.118 &  0.029 &  0.089 \\
Tol0505-387  & in & -0.086 &  0.036 &  0.037 &  0.135 &  0.038 &  0.043 \\
Tol0510-400  & in & -0.352 & -0.082 & -0.049 &  0.047 &  0.071 &  0.185 \\
Tol0528-383(W)  & re & -0.771 &  0.118 & -0.118 &  0.051 &  0.002 &  0.090 \\
Tol0528-383(E)  & re & -0.575 &  0.158 & -0.022 &  0.032 & -0.021 &  0.056 \\
Tol0528-383  & in & -0.476 &  0.156 &  0.085 &  0.084 & -0.001 &  0.009 \\
Tol0538-416  & in & -0.359 & -0.081 & -0.019 &  0.004 &  0.102 &  0.033 \\
Tol0610-387  & in & -0.698 & -0.038 &  0.132 &  0.158 & -0.040 & -0.019 \\
Tol0633-415  & in & -0.582 & -0.043 &  0.511 &  0.020 &  0.322 &  0.145 \\
Tol0645-376  & in & -0.474 &  0.073 &  0.092 &  0.068 &  0.042 &  0.028 \\
Tol0957-278(NW)  & re & -0.582 &  0.010 &  0.024 &  0.012 &  0.029 &  0.048 \\
Tol0957-278(SE)  & re & -0.631 &  0.015 &  0.035 &  0.031 &  0.010 &  0.033 \\
Tol0957-278  & in & -0.651 &  0.019 &  0.038 &  0.038 &  0.005 &  0.035 \\
Tol1004-296(NW)  & re & -0.538 &  0.067 &  0.057 &  0.030 &  0.018 &  0.052 \\
Tol1004-296(SE)  & re & -0.594 &  0.046 &  0.045 &  0.022 &  0.019 &  0.046 \\
Tol1004-296  & in & -1.895 & -0.060 &  0.032 &  0.004 &  0.027 &  0.036 \\\hline

\end{tabular}
\end{table*}
\addtocounter{table}{-1}
   \begin{table*}
      \caption{Indices of individual spectra (continued).}
         \begin{tabular}{lrrrrrrr}\hline
            
galaxy & type & cont. i. & D(4000) & H+K(Ca) & H$_{\delta}$ & Fe4531 & Mg$_b$ \\\hline

Tol1008-286(SE)  & re & -0.261 &  0.449 & -0.062 & -0.227 &  0.026 & -0.189 \\
Tol1008-286(NW)  & re & -0.974 & -0.037 &  0.046 & -0.048 & -0.122 &  0.439 \\
Tol1008-286  & in & -0.369 &  0.150 &  0.149 & -0.670 &  0.178 &  0.071 \\
Tol1025-285  & in & -0.589 &  0.083 &  0.168 &  0.002 & -0.021 &  0.148 \\
Tol1146-333  & in & -0.419 &  0.178 &  0.147 &  0.073 &  0.100 & -0.014 \\
Tol1147-283  & in & -0.460 &  0.062 &  0.161 &  0.047 &  0.098 &  0.127 \\
Tol1223-359  & in & -0.723 & -0.105 & -0.194 &  0.000 &  0.026 & -0.372 \\
Tol1345-420  & in & -0.430 &  0.119 &  0.080 &  0.065 &  0.038 &  0.073 \\
Tol1455-284(E)  & re & -0.121 & -0.036 & -0.012 &  0.082 & -0.054 & -0.041 \\
Tol1455-284(W)  & re &  0.126 &  0.152 &  0.023 &  0.072 & -0.006 &  0.076 \\
Tol1455-284  & in &  0.102 &  0.130 &  0.144 & -0.013 &  0.012 &  0.094 \\
Tol1457-262E(NW)     & re & -0.484 &  0.026 &  0.060 &  0.007 &  0.024 &  0.079 \\
Tol1457-262E(SE)     & re & -0.503 &  0.021 &  0.063 &  0.007 &  0.029 &  0.074 \\
Tol1457-262E         & in & -0.536 &  0.011 &  0.034 & -0.005 &  0.052 &  0.076 \\
Tol1457-262W(W)      & re & -0.553 &  0.001 &  0.086 &  0.048 & -0.015 &  0.021 \\
Tol1457-262W(CentW)  & re & -0.571 &  0.016 &  0.096 &  0.049 &  0.008 &  0.051 \\
Tol1457-262W(Cent)   & re & -0.472 &  0.082 &  0.119 &  0.071 &  0.061 &  0.056 \\
Tol1457-262W(CentE)  & re & -0.513 &  0.055 &  0.103 &  0.063 &  0.026 &  0.050 \\
Tol1457-262W(E)  & re & -0.558 &  0.030 &  0.096 &  0.097 & -0.041 &  0.011 \\
Tol1457-262W  & in & -0.145 &  0.315 & -0.062 & -0.013 & -0.009 &  0.037 \\
Tol1924-416(W)  & re & -0.656 & -0.018 &  0.043 & -0.010 &  0.027 &  0.045 \\
Tol1924-416(E)  & re & -0.761 & -0.013 &  0.068 & -0.018 &  0.026 &  0.051 \\
Tol1924-416  & in & -0.679 & -0.016 &  0.054 & -0.014 &  0.027 &  0.048 \\
Tol1937-423  & in & -0.596 &  0.092 &  0.209 &  0.075 &  0.039 &  0.047 \\
Tol2019-405  & in & -0.382 &  0.128 &  0.045 &  0.025 &  0.042 &  0.017 \\
Tol2122-408  & in & -0.410 &  0.183 & -0.017 &  0.062 &  0.029 &  0.057 \\
Tol2138-397  & in & -0.554 &  0.177 &  0.221 & -0.038 &  0.090 &  0.097 \\
Tol2146-391  & in & -0.725 &  0.258 &  0.045 & -0.090 & -0.270 & -0.215 \\
Tol2240-384  & in & -0.360 & -0.059 & -0.141 &  0.035 & -0.100 & -0.149 \\
UM69(E)      & re & -1.314 &  0.020 &  0.112 &  0.056 &  0.044 &  0.029 \\
UM69(Cent)   & re & -0.650 &  0.077 &  0.122 &  0.066 &  0.047 &  0.047 \\
UM69(W)      & re & -0.744 & -0.026 &  0.086 &  0.052 &  0.051 &  0.050 \\
UM69         & in & -1.058 &  0.013 &  0.108 &  0.058 &  0.047 &  0.041 \\
UM137(E)     & re & -1.467 &  0.086 &  0.036 &  0.150 &  0.069 &  0.044 \\
UM137(CentE) & re & -0.619 &  0.148 &  0.104 &  0.107 &  0.033 &  0.047 \\
UM137(CentW) & re & -0.587 &  0.123 &  0.085 &  0.116 &  0.035 &  0.039 \\
UM137(W)     & re & -0.423 &  0.041 &  0.060 &  0.137 &  0.053 &  0.020 \\
UM137        & in & -0.462 &  0.042 &  0.071 &  0.142 &  0.016 &  0.045 \\
UM151        & in & -0.482 &  0.084 &  0.547 & -0.084 &  0.073 &  0.067 \\
UM160(E)     & re & -0.709 &  0.019 &  0.106 &  0.014 &  0.141 &  0.238 \\
UM160(Cent)  & re & -0.865 & -0.051 &  0.071 &  0.039 &  0.066 &  0.165 \\
UM160(W)     & re & -0.633 & -0.012 &  0.131 &  0.059 &  0.016 &  0.104 \\
UM160        & in & -0.690 & -0.032 &  0.114 &  0.019 &  0.038 &  0.171 \\
UM166        & in & -0.460 &  0.242 &  0.182 &  0.041 &  0.062 &  0.146 \\
UM191        & in & -0.593 &  0.102 &  0.132 &  0.057 &  0.039 &  0.073 \\
UM238(E)     & re & -0.584 &  0.154 &  0.149 &  0.030 &  0.024 &  0.090 \\
UM238(W)     & re & -0.177 &  0.147 &  0.103 & -0.005 &  0.121 &  0.021 \\
UM238        & in & -0.518 &  0.139 &  0.139 &  0.032 &  0.057 &  0.060 \\
UM306        & in & -0.520 &  0.233 &  0.145 &  0.067 &  0.026 &  0.041 \\
UM307        & in & -0.709 &  0.071 &  0.125 &  0.059 &  0.057 &  0.085 \\\hline

\end{tabular}
\end{table*}
\addtocounter{table}{-1}
   \begin{table*}
      \caption{Indices of individual spectra (continued).}
         \begin{tabular}{lrrrrrrr}\hline
galaxy & type & cont. i. & D(4000) & H+K(Ca) & H$_{\delta}$ & Fe4531 & Mg$_b$ \\\hline

UM323        & in & -1.140 &  0.020 &  0.046 &  0.085 &  0.040 &  0.038 \\
UM382        & in & -0.512 &  0.014 &  0.123 &  0.207 &  0.037 & -0.191 \\
UM391        & in & -0.158 &  0.094 & -0.241 & -0.030 & -0.029 &  0.001 \\
UM395        & in & -0.651 &  0.125 &  0.132 &  0.100 &  0.059 &  0.081 \\
UM396        & in & -0.665 &  0.017 &  0.000 & -0.001 &  0.038 & -0.082 \\
UM408        & in & -0.632 &  0.105 & -0.152 & -0.002 &  0.046 &  0.042 \\
UM417        & in & -0.550 &  0.126 &  0.438 & -0.024 &  0.175 & -0.173 \\
UM439(SE)    & re & -0.410 &  0.100 &  0.136 & -0.044 &  0.045 &  0.024 \\
UM439(Cent)  & re & -0.585 &  0.052 &  0.103 &  0.073 &  0.040 &  0.046 \\
UM439(NW)    & re & -0.579 &  0.053 &  0.106 &  0.043 &  0.051 &  0.021 \\
UM439        & in & -0.634 &  0.036 &  0.093 &  0.019 &  0.048 &  0.003 \\
UM448        & in & -0.470 &  0.104 &  0.125 &  0.034 &  0.036 &  0.098 \\
UM448(extension NE-SW)  & in & -0.504 &  0.065 &  0.094 &  0.012 &  0.040 &  0.119 \\
UM455(NW)    & re & -0.469 &  0.102 &  0.067 & -0.020 &  0.026 & -0.018 \\
UM455(SE)    & re & -0.116 &  0.049 &  0.168 & -0.071 &  0.098 & -0.029 \\
UM455        & in &  0.465 &  0.005 &  0.417 & -0.034 &  0.060 & -0.003 \\
UM456(NE)    & re & -0.742 & -0.011 &  0.108 &  0.021 &  0.101 &  0.041 \\
UM456(Cent)  & re & -0.754 & -0.012 &  0.068 &  0.027 &  0.069 &  0.036 \\
UM456(SW)    & re & -0.750 & -0.034 &  0.048 &  0.029 &  0.024 & -0.002 \\
UM456        & in & -0.732 &  0.021 &  0.080 &  0.026 &  0.047 &  0.064 \\
UM461(E)     & re & -0.686 &  0.061 & -0.001 & -0.091 & -0.087 &  0.047 \\
UM461(W)     & re & -0.728 &  0.043 &  0.053 & -0.084 & -0.100 &  0.034 \\
UM461        & in & -0.808 & -0.007 &  0.182 & -0.066 & -0.036 &  0.001 \\
UM462(SW)    & re & -0.884 & -0.048 &  0.002 &  0.011 &  0.040 &  0.028 \\
UM462(NE)    & re & -0.755 & -0.017 &  0.019 &  0.014 &  0.031 &  0.085 \\
UM462        & in & -0.796 & -0.024 &  0.004 &  0.014 &  0.036 &  0.058 \\
UM463        & in & -0.669 & -0.764 & -0.243 & -0.058 &  0.069 &  0.257 \\
UM477        & in & -0.229 &  0.202 &  0.160 &  0.017 &  0.046 &  0.107 \\
UM483        & in & -0.482 &  0.078 &  0.020 &  0.078 &  0.000 &  0.123 \\
UM499(W)     & re & -0.569 &  0.080 &  0.102 &  0.088 &  0.039 &  0.095 \\
UM499(E)     & re & -0.447 &  0.155 &  0.131 &  0.019 &  0.045 &  0.083 \\
UM499        & in & -0.437 &  0.155 &  0.136 &  0.023 &  0.046 &  0.087 \\
UM533(E)     & re & -0.655 &  0.100 &  0.097 &  0.019 &  0.042 &  0.108 \\
UM533(W)     & re & -0.535 &  0.004 & -0.016 & -0.019 &  0.055 &  0.034 \\
UM533        & in & -0.482 &  0.058 &  0.006 & -0.025 &  0.046 &  0.044 \\
UM559(giant E-RHII)  & re & -0.784 &  0.073 & -0.260 & -0.088 & -0.115 &  0.392 \\
UM559(E)     & re & -0.638 & -0.079 & -0.586 &  0.210 & -0.975 &  0.959 \\
UM559(Cent)  & re & -0.683 &  0.158 &  0.240 &  0.047 &  0.087 &  0.118 \\
UM559        & in & -0.459 &  0.154 &  0.139 & -0.211 &  0.307 &  0.381 \\
UM570        & in & -0.674 &  0.464 &  1.274 & -0.274 &  0.732 & -1.148 \\
UM598E(SW)   & re & -0.217 &  0.147 &  0.079 &  0.117 &  0.014 &  0.015 \\
UM598E(Cent) & re & -0.485 &  0.291 &  0.274 &  0.013 &  0.057 &  0.072 \\
UM598E(NE)   & re & -0.935 &  0.169 &  0.184 &  0.050 &  0.047 &  0.054 \\
UM598E       & in & -0.858 &  0.106 &  0.172 &  0.056 &  0.054 &  0.043 \\
UM598W(SW)   & re & -0.578 &  0.080 &  0.069 &  0.007 &  0.051 &  0.084 \\
UM598W(CentSW)  & re & -0.339 &  0.145 &  0.163 &  0.061 &  0.041 &  0.101 \\
UM598W(CentNE)  & re & -0.392 &  0.264 &  0.197 &  0.015 &  0.060 &  0.130 \\
UM598W(NE)   & re & -0.473 &  0.140 &  0.130 &  0.040 &  0.042 &  0.086 \\
UM598W       & in & -0.574 &  0.103 &  0.119 &  0.042 &  0.018 &  0.049 \\
IIZw40       & in & -0.363 & -0.024 &  0.017 & -0.055 &  0.020 &  0.059 \\\hline
            
         \end{tabular}
   \end{table*}

   \begin{table*}
   \begin{center}
      \caption{Population parameters of individual spectra.}
      \label{solutions}
         \begin{tabular}{lrrrrr}
            \hline
galaxy  & ${\rm [Fe/H]}_{y+i}$ & \multicolumn{2}{c}{``BC99''}  & \multicolumn{2}{c}{``Starburst''}  \\
  & & $M_{y+i}$:$M_{o}$ & $age_{y+i}$[Myr] & $M_{y+i}$:$M_{o}$ & $age_{y+i}$[Myr] \\
            \hline
Cam08-28A(Cent)  &     -1 &   1:10 &    102 &   1:10 &    100 \\
Cam08-28A(E)  &     -1 &  1:100 &     10 &  1:100 &     10 \\
Cam08-28A    & -0.744 &  1:100 &     10 &  1:100 &     10 \\
Cam0840+1044  & -1.140 &    1:3 &    102 &    1:3 &    100 \\
Cam0840+1201  & -0.894 &   1:30 &     10 &  1:100 & 2 \\
Cam1543+0907  & -1.163 &  1:100 & 2 &  1:100 & 1 \\
CTS1008      & -0.713 &   1:30 & 1 &   1:30 & 1 \\
CTS1011      & -0.692 &  1:100 & 2 &  1:100 & 2 \\
CTS1016      &     -1 &  1:100 &     10 &  1:100 &     20 \\
CTS1018      & -0.900 &  1:100 &     10 &  1:100 &     20 \\
CTS1019      & -0.646 &  1:100 & 2 &  1:100 & 1 \\
CTS1022      & -0.781 &  1:100 & 1 &  1:100 &     10 \\
CTS1028      & -0.820 &   1:30 & 2 &   1:30 & 2 \\
CTS1029      &     -1 &  1:100 & 2 &  1:100 &     10 \\
CTS1033      & -0.862 &  1:100 &    102 &   1:30 &    200 \\
CTS1034      & -0.907 &  1:100 &     10 &  1:100 &     20 \\
DDO060(NW)   &     -1 &   1:30 &    203 &   1:30 &    200 \\
DDO060(SE)   &     -1 &   1:10 &    509 &   1:10 &    500 \\
DDO070(E)    & -1.410 &  1:100 &     10 &  1:100 &     20 \\
DDO075(CentSW) &   -1 &  1:100 &      5 &  1:100 &      5 \\
DDO155(SW)   & -1.229 &  1:100 & 5 &  1:100 &     10 \\
DDO155       & -1.109 &    1:0 &     20 &    1:1 & 5 \\
ESO289IG037  &     -1 &  1:100 &    203 &  1:100 &    100 \\
Fairall30    &     -1 &  1:100 &     20 &  1:100 &     20 \\
Haro24       &     -1 &   1:30 &     50 &  1:100 &     20 \\
Haro25       &     -1 &  1:100 &     50 &  1:100 &     50 \\
Haro30       & -1.201 &  1:100 &     50 &  1:100 &     50 \\
IC5154(N)    &     -1 &  1:100 &     50 &   1:30 &    100 \\
Marseille68  &     -1 &  1:100 &     50 &  1:100 &     50 \\
Marseille88  & -0.851 &    1:3 &     50 &    1:3 &     50 \\
MBG20533-4410  &     -1 &  1:100 & 1 &  1:100 & 1 \\
MBG21567-1645  &     -1 &  1:100 &    102 &  1:100 &    100 \\
MBG22012-1550(E)  &     -1 &   1:30 &     50 &   1:30 &     50 \\
MCG0157017   &     -1 &  1:100 &     10 &  1:100 &     20 \\
Mrk36        & -0.976 &  1:100 & 5 &  1:100 & 2 \\
Mrk710(CentNE)  &     -1 &   1:10 &     50 &  1:100 &     10 \\
Mrk710(NE)   &     -1 &   1:30 &     20 &  1:100 & 2 \\
Mrk710       &     -1 &  1:100 &     10 &  1:100 &     10 \\
Mrk1318      & -0.598 &  1:100 &     50 &   1:30 &     50 \\
NGC7323(E)   &     -1 &   1:30 &    102 &   1:30 &    100 \\
NGC7323(W)   &     -1 &  1:100 &     50 &   1:30 &    100 \\
            \hline
         \end{tabular}
   \end{center}
   \end{table*}
\addtocounter{table}{-1}
   \begin{table*}
   \begin{center}
      \caption{Population parameters of individual spectra (continued).}
         \begin{tabular}{lrrrrr}
            \hline
galaxy  & ${\rm [Fe/H]}_{y+i}$ & \multicolumn{2}{c}{``BC99''}  & \multicolumn{2}{c}{``Starburst''}  \\
  & & $M_{y+i}$:$M_{o}$ & $age_{y+i}$[Myr] & $M_{y+i}$:$M_{o}$ & $age_{y+i}$[Myr] \\
            \hline
Tol0104-388(NW)  & -0.908 &   1:30 & 2 &   1:10 & 1 \\
Tol0117-414EW  &     -1 &   1:30 &     50 &  1:100 &     20 \\
Tol0117-414NS(N)  &     -1 &  1:100 & 1 &  1:100 &     10 \\
Tol0117-414NS(CentS)  &     -1 &  1:100 &     10 &  1:100 &     20 \\
Tol0140-420  & -0.812 &  1:100 & 1 &  1:100 &     10 \\
Tol0226-390  & -0.725 &  1:100 & 2 &  1:100 & 1 \\
Tol0306-405  & -0.728 &  1:100 & 1 &  1:100 &     10 \\
Tol0341-407(E)  & -0.738 &   1:30 &     20 &  1:100 &     10 \\
Tol0341-407(W)  & -0.755 &   1:30 &     20 &  1:100 & 1 \\
Tol0440-381  & -0.906 &    1:3 &     50 &   1:30 &     20 \\
Tol0510-400  &     -1 &   1:30 &     10 &   1:30 & 1 \\
Tol0528-383(W)  & -0.863 &    1:3 &    203 &    1:3 &    200 \\
Tol0528-383(E)  & -0.722 &   1:30 &    102 &   1:30 &    100 \\
Tol0528-383  & -0.912 &   1:30 &    203 &   1:30 &    200 \\
Tol0538-416  & -0.870 &  1:100 & 2 &  1:100 & 2 \\
Tol0610-387  &     -1 &    1:1 &     50 &    1:1 &     50 \\
Tol0645-376  & -1.100 &  1:100 &     20 &  1:100 &     20 \\
Tol0957-278(NW)  & -0.761 &  1:100 & 1 &  1:100 & 1 \\
Tol0957-278(SE)  & -0.972 &  1:100 & 1 &  1:100 &     10 \\
Tol0957-278  & -0.846 &  1:100 & 1 &  1:100 &     10 \\
Tol1004-296(NW)  & -0.592 &  1:100 &     20 &  1:100 &     20 \\
Tol1004-296(SE)  & -0.666 &  1:100 &     10 &  1:100 &     10 \\
Tol1008-286(NW)  &     -1 &   1:30 &      1 &   1:10 &      2 \\
Tol1147-283  & -0.992 &  1:100 &     20 &  1:100 &     20 \\
Tol1223-359  &     -1 &   1:30 & 2 &    1:0 & 1 \\
Tol1345-420  & -0.610 &   1:30 &    102 &   1:30 &     50 \\
Tol1457-262E(NW)  & -0.653 &  1:100 &     10 &  1:100 &     10 \\
Tol1457-262E(SE)  & -0.805 &  1:100 & 1 &  1:100 &     10 \\
Tol1457-262E  & -0.705 &  1:100 & 1 &  1:100 & 1 \\
Tol1457-262W(W)  &     -1 &  1:100 & 1 &  1:100 &     10 \\
Tol1457-262W(CentW)  & -0.903 &  1:100 & 1 &  1:100 &     10 \\
Tol1457-262W(Cent)  & -0.737 &  1:100 &     20 &   1:30 &     50 \\
Tol1457-262W(CentE)  & -0.787 &   1:30 &     50 &  1:100 &     20 \\
Tol1457-262W  & -0.748 &  1:100 &    509 &  1:100 &    500 \\
Tol1924-416(W)  & -0.855 &  1:100 & 2 &  1:100 & 2 \\
Tol1924-416(E)  & -0.874 &  1:100 & 5 &  1:100 & 2 \\
Tol1924-416  & -0.863 &  1:100 & 2 &  1:100 & 2 \\
Tol1937-423  &     -1 &   1:30 &     50 &   1:30 &     50 \\
Tol2019-405  & -0.882 &  1:100 &     50 &   1:30 &    100 \\
Tol2122-408  &     -1 &   1:10 &    509 &   1:10 &    500 \\
            \hline
         \end{tabular}
   \end{center}
   \end{table*}
\addtocounter{table}{-1}
   \begin{table*}
   \begin{center}
      \caption{Population parameters of individual spectra (continued).}
         \begin{tabular}{lrrrrr}
            \hline
galaxy  & ${\rm [Fe/H]}_{y+i}$ & \multicolumn{2}{c}{``BC99''}  & \multicolumn{2}{c}{``Starburst''}  \\
  & & $M_{y+i}$:$M_{o}$ & $age_{y+i}$[Myr] & $M_{y+i}$:$M_{o}$ & $age_{y+i}$[Myr] \\
            \hline
UM69(E)      & -0.818 &  1:100 & 5 &   1:30 & 5 \\
UM69(Cent)   & -1.147 &  1:100 &     20 &  1:100 &     20 \\
UM69(W)      &     -1 &  1:100 & 5 &  1:100 & 5 \\
UM69         & -0.672 &  1:100 & 5 &   1:30 & 5 \\
UM137(CentE)  &     -1 &   1:30 &    203 &   1:10 &    500 \\
UM137(CentW)  &     -1 &   1:10 &    203 &   1:10 &    200 \\
UM137(W)     &     -1 &   1:10 &    102 &   1:10 &    100 \\
UM160(E)     &     -1 &  1:100 & 5 &  1:100 &     10 \\
UM160(Cent)  &     -1 &   1:30 & 1 &   1:30 &     10 \\
UM160(W)     &     -1 &  1:100 & 2 &  1:100 &     10 \\
UM160        &     -1 &  1:100 & 5 &  1:100 & 2 \\
UM191        &     -1 &  1:100 &     50 &   1:30 &    100 \\
UM323        & -0.950 &    1:0 &     50 &   1:30 & 5 \\
UM395        &     -1 &   1:30 &    203 &   1:30 &    100 \\
UM396        & -0.688 &  1:100 & 2 &  1:100 & 1 \\
UM408        & -0.846 &   1:10 &    102 &   1:10 &    100 \\
UM439(Cent)  & -1.076 &   1:30 &     50 &  1:100 &     20 \\
UM439(NW)    &     -1 &  1:100 &     10 &  1:100 &     10 \\
UM439        & -0.861 &  1:100 &     10 &  1:100 &     10 \\
UM448        & -0.866 &  1:100 &     50 &  1:100 &     50 \\
UM448(extension NE-SW)  & -0.704 &  1:100 &     20 &  1:100 &     20 \\
UM456(NE)    & -0.912 &  1:100 & 5 &  1:100 & 5 \\
UM456(Cent)  & -0.817 &  1:100 & 5 &  1:100 & 5 \\
UM456(SW)    & -0.923 &  1:100 & 5 &   1:30 &     10 \\
UM456        & -0.922 &  1:100 & 1 &  1:100 &     10 \\
UM461(E)     & -1.068 &  1:100 & 1 &  1:100 & 1 \\
UM461(W)     & -1.330 &  1:100 & 5 &  1:100 & 2 \\
UM461        & -1.101 &  1:100 & 5 &  1:100 & 2 \\
UM462(SW)    & -0.962 &   1:30 & 1 &   1:30 & 2 \\
UM462(NE)    & -0.914 &  1:100 & 5 &  1:100 & 2 \\
UM462        & -0.926 &  1:100 & 5 &   1:30 & 1 \\
UM477        &     -1 &  1:100 &    509 &  1:100 &    500 \\
UM483        &     -1 &   1:30 &     50 &   1:30 &     50 \\
UM499(W)     &     -1 &   1:30 &    102 &   1:30 &     50 \\
UM533(E)     & -0.503 &  1:100 &     20 &  1:100 &     20 \\
UM533(W)     &     -1 &  1:100 & 2 &  1:100 & 1 \\
UM533        & -0.598 &  1:100 &     10 &  1:100 &     10 \\
UM598E(SW)   &     -1 &   1:10 &    509 &   1:10 &    500 \\
UM598W(SW)   &     -1 &  1:100 &     20 &  1:100 &     20 \\
UM598W(CentSW)  &     -1 &  1:100 &    102 &  1:100 &     50 \\
UM598W(NE)   &     -1 &  1:100 &     50 &  1:100 &     50 \\
IIZw40       & -0.796 &  1:100 & 2 &  1:100 & 2 \\
            \hline
         \end{tabular}
   \end{center}
   \end{table*}

\end{document}